\documentclass{aa}  

\usepackage{graphicx}
\usepackage{txfonts}
\usepackage{bm}
\usepackage{multirow}
\usepackage{mathtools}
\usepackage[referable]{threeparttablex}
\usepackage{lscape}
\begin{document}

   \title{Internal Structure and Magnetic Moment of Rocky Planets}

   \subtitle{Application to the first exoplanets discovered by TESS}

   \author{J.M. Rodr\'{\i}guez-Mozos
          \inst{1}
          \and
          A. Moya\inst{1,2}
          }

   \institute{
            Departament d’Astronomia i Astrofísica, Universitat de València, C. Dr. Moliner 50, 46100 Burjassot, Spain
\and
             Electrical Engineering, Electronics, Automation and Applied Physics Department, E.T.S.I.D.I, Polytechnic University of Madrid (UPM), Madrid 28012, Spain\\
             \email{andres.moya-bedon@uv.es}
             }

   \date{Received November 29, 2021; accepted March 16, 1997}

 
  \abstract
   {For a planet to be considered habitable on its surface, it is an important advantage for it to have a magnetic field that protects its atmosphere from stellar winds as well as cosmic rays. Magnetic protection of potentially habitable planets plays a key role in determining the chances of detecting atmospheric biosignatures. In making an estimate of a planet’s magnetic field, its internal structure needs to be known first.}
   {This paper proposes to use the Preliminary Reference Earth Model (PREM) internal structure as the base of a numerical model. PREM is considered the best available option for estimating the internal structure of rocky exoplanets. With this model, we estimate the magnetic properties of dry and water-rich Earth-like and Super-Earth-like planets. We apply it to those of this kind at the first 176 planets confirmed by the TESS exoplanet mission.}
   {Using PREM as a reference, we estimate the internal structure of dry and water-rich rocky planets. This model provides an estimation of the average density and core size of the planet. These are the key ingredients for estimating its magnetic moment depending on whether it is tidally locked or not. Our model estimates the thermodynamic variables as a function of pressure, and including saltwater as a component of water-rich exoplanets. In addition, we have not used the perfect layer differentiation approximation. We have validated our model with those planets and satellites in the Solar system with similar characteristics.}
   {Using our model, we have estimated the internal structure and magnetic moment of those dry and water-rich rocky planets and satellites in the Solar system. The differences with the observed values in the internal structure characteristics, mass, average density, moment of inertia factor, and local Rossby number are remarkably low or even negligible. The estimated magnetic moments are also very similar to the observed ones. We have applied the model to the first 176 planets confirmed by the TESS, finding that, from an astrobiological perspective TOI-700 d and TOI-2257 b are the most interesting ones as being located in the habitable zone (HZ), although their magnetic moments are only about 0.01 of the Earth's magnetic moment.
}


   \maketitle
%

\section{Introduction}

For a planet to be considered habitable on its surface, it is generally advantageous for it to have a magnetic field that protects its atmosphere from stellar winds as well as cosmic rays. Magnetic protection of potentially habitable planets plays a key role in determining their habitability and the chances of detecting biomarkers in its atmosphere \citep{Zuluaga13}. An Earth-like planet without a magnetic field orbiting an active low-mass star can lose its ozone column drastically in a relatively short time \citep{Tilley17}. If this happens, high-energy cosmic rays can reach the planet's surface significantly altering the development of life.

Planetary dynamos are generally thought to be maintained by thermal and compositional convection mechanisms in electrically conducting fluids located in the planet interiors \citep[][OC06]{Olson06}. The magnetic field generated by the dynamo of a rocky planet depends mainly on the density and size of the convective core, and the convective buoyancy flux generated in this core. Therefore, for estimating the magnetic field generated by a rocky planet, its internal structure needs to be known first. Unfortunately, the interior structure of an exoplanet is hidden from direct observation. For estimating its internal structure from only the Mass and Radius of the exoplanet, we are limited to indirect methods based on theoretical models \citep{Suissa18}. 

Current technological state-of-the-art provides, in the best scenario, planetary masses, radii, and orbital periods with uncertainties usually larger than a 10$\%$. In many other cases, their masses or radii are unknown and we must use models for estimating them \citep{Chen17}. In this uncertain context, our approximation to solve this problem of describing the internal structure of an exoplanet is to extrapolate it from the internal structure of known objects at the solar system.

In the case of rocky exoplanets, extrapolation from the solar system to know their internal composition implies that a minimum of three primary constituents must be used: iron, enstatite (Mg SiO$_3$), and water \citep{Valencia06}. In general, for a given chemical composition, the Equation Of State (EOS) of any material can be expressed as:
\begin{equation}
    \rho=\rho(P,T)
\end{equation}

\noindent where $\rho$ is the density, $P$ the pressure, and $T$ the temperature.

The effect of temperature is secondary compared to the effect of pressure for denser constituents such as iron or enstatite \citep{Valencia06}. For lighter elements, such as water, the effect of temperature cannot be disregarded but a pressure-temperature relation such as the water melting curve can be included in the models for taking it into account \citep{Zeng13}.  

Dry rocky planets are those whose water mass is negligible with respect to the total mass of the planet, and therefore its internal structure can be explained by only two components. Nevertheless, the core may also be affected by eutectic melting, most likely due to sulfur. This would be the case for the inner planets of the solar system such as Mercury, Venus, Earth, and Mars. When there are only two basic constituents, for each mass and radius value of the planet the theoretical models provide a unique solution for its internal structure \citep{Suissa18, Zeng13}.

However, water-rich rocky exoplanets need three basic components to explain their internal structure: iron in the core, enstatite in the mantle, and water usually forming an outer layer of ice. In this case, for each mass and radius value, the theoretical models provide infinite internal structure solutions all compatible with those mass and radius values \citep{Suissa18, Zeng13}.  

One of the most used assumptions for modelling these different layers is their complete differentiation. Theoretical models using equations of state of pure components need the layers to be completely differentiated, i.e. all the iron on the planet must be in the nucleus and all the silicates in the mantle. This complete differentiation does not occur on the inner planets of the solar system and, most likely, it will be difficult to find completely differentiated exoplanets.

Currently, we have the PREM \citep{Dzie81} which, through a wide and extensive seismic field study, has allowed us to determine the physical characteristics of the different constituent layers of our planet. The PREM model provides us, apart from many other data, a realistic relationship between density and pressure inside Earth, despite the presence of impurities and partial lack of differentiation.

On the other hand, recent studies on the spectrum of white dwarfs contaminated with the remnants of disintegrated planets show that these remnants are quite similar to the earth's composition. More than 85$\%$ of the mass of these remnants are Fe, Mg, Si, and O. In addition, they have Fe/Si and Mg/Si ratios similar to terrestrial ones, and they are accompanied by poor C content. This leads us to assume that these disintegrated planets had a formation and evolution process similar to the inner planets of the solar system \citep{Jura14}.

In this paper, we aim to construct an internal structure model for rocky planets. The model is a PREM-based internal structure model that is considered the best available option for estimating the internal structure of these rocky exoplanets, extrapolating the behaviour of density within Earth following PREM and performing only those simplifications that are considered strictly necessary. Regarded to water-rich rocky planets, we have extrapolated the internal structure of the cases of the water-rich Jupiter's moons (Europa, Ganymede, and Calixto).

With this estimated internal structure, we have analysed the temperature profile and the heat crossing the boundary between the planetary nucleus and mantle. The main idea is to use the planetary dynamo scaling laws (OC06) for estimating the magnetic regime (dipolar/multipolar), moment, and field. In summary, we estimate the planetary magnetic shield protecting its atmosphere from erosion provoked by the stellar wind and cosmic rays \citep{Mozos19}. Finally, we have applied this model to all the confirmed rocky planets (dry and water-rich) discovered by TESS \citep{TESS}.

\section{Internal structure of dry rocky planets}

\begin{figure}
   \centering
    \includegraphics{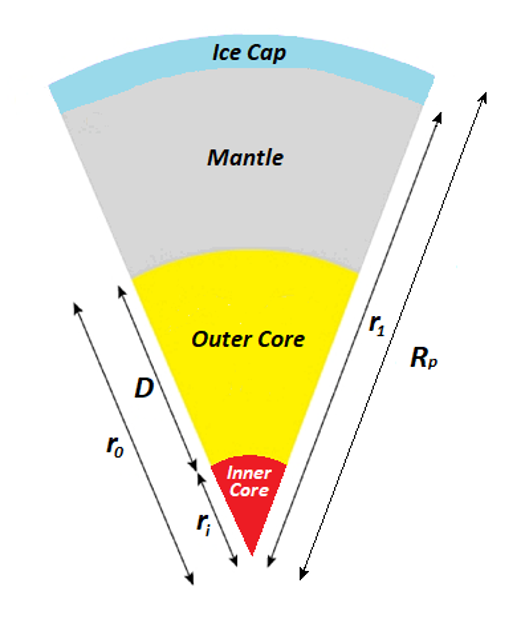}
   \caption{Section of a rocky planet. $R_p$ is the Planet radius, $r_0$ the core radius, $r_1$ the Mantle radius, $D$ the Outer core thickness, and $r_i$ the Inner core radius. 
}
              \label{Estruct_int}%
    \end{figure}

Consider a rocky planet from which we know its normalized mass ($M_p=\frac{M}{M_\oplus}$) and its normalized radius ($R_p=\frac{R}{R_\oplus}$) relative to Earth values. In general, this type of planet will have three layers as can be seen in a cross-section (Fig. \ref{Estruct_int}). In this figure, we can see the innermost layer or core of the planet, that it is basically made up of Fe, the mantle whose basic components are silicates and oxides of Si, Mg, and Fe, and finally an outer layer usually of water ice where liquid internal oceans can also exist. When the ice cap does not exist or its mass is negligible compared to the total mass of the planet, it will be considered a dry rocky planet. In this case, the radius of the mantle will be the radius of the planet ($r_1 \approx R_p$) and the internal structure of the planet will be defined by only two layers whose basic elements will be the normalized radius of the core ($r_0$), the normalized average core density ($\rho_0$) and the normalized average density of the mantle ($\rho_1$). For obtaining these values it would be necessary to determine the planetary density and pressure profiles with the radius.

On Earth, however, the planet's core consists of two distinct areas. On the one hand, the outer core is dominated by liquid Fe, with a few percent of Ni and 5-15$\%$ of lighter elements we present in following sections, whose thickness we will call $D$, and the inner core, which is composed by crystalline solid Fe, on the other. In general, for a dry rocky exoplanet, of which only $M_p$ and $R_p$ are known, it would be necessary to use models for the thermal evolution of the planet to determine if part of the nucleus is in solid-state and the size of this zone.

\subsection{Internal composition of the Earth}

The Earth's interior is basically composed, as we have already said, of four chemical elements: Fe, Mg, Si, and O. The core consists mainly of Fe and in much smaller quantity by other lighter elements such as S, Si, O, and C \citep{Fischer12}. Nevertheless, there is not a consensus in the literature about which elements are part of the list of these secondary components, and in which percentage.

Above 100 GPa, Iron is mostly in phase $\epsilon$ with a hexagonal crystalline structure \citep{Zeng13}. Given that the pressure at the core-mantle boundary (CMB) of the Earth reaches 135 GPa, the solid Fe contained in the core will be all in phase $\epsilon$.

Regarding the mantle, it consists mainly of Mg, Si, O, and to a lesser extent Al, Ca, and Na \citep{Sotin07}. The most abundant compounds in the Earth's mantle are magnesium silicates, such as enstatite (Mg Si O$_3$) and olivine ((Fe,Mg)$_2$ Si O$_4$), and oxides like periclase (Mg O) and silica (Si O$_2$). Magnesium silicates often incorporate impurities of Fe that, in the case of enstatite, can reach up to 12$\%$. As the Mg/Si ratio increases, the relative quantity of olivine and periclase also increases. Above 27 Gpa, the polymorphs of olivine are unstable and, then, the terrestrial mantle is basically composed of enstatite polymorphs such as perovskite (pv) and post-perovskite (ppv), and periclase, being this last element only a 7 $\%$ of the total, approximately. Reaching 125 GPa and 2500 K, the pv is transformed into ppv with a density jump of around 1.5$\%$. Around 900 GPa there is a dissociation of ppv into periclase and compact silicates (Mg Si$_2$ O$_5$). Above 2100 Gpa there is the second dissociation of ppv into periclase and silica \citep{Zeng13}.

\subsection{PREM}

\citet{Dzie81} presented an internal structure model for the Earth obtained using seismic data (PREM). The velocity of the seismic waves at different points is a reflection of the physical characteristics of the different zones they have traveled by. PREM can be considered to be the best approximation up to now to the Earth's internal structure and it has accurately determined the Earth's mass and radius, as well as the pressure and density as a function of the Earth's radius.

\begin{figure}
   \centering
    \includegraphics[width=\linewidth]{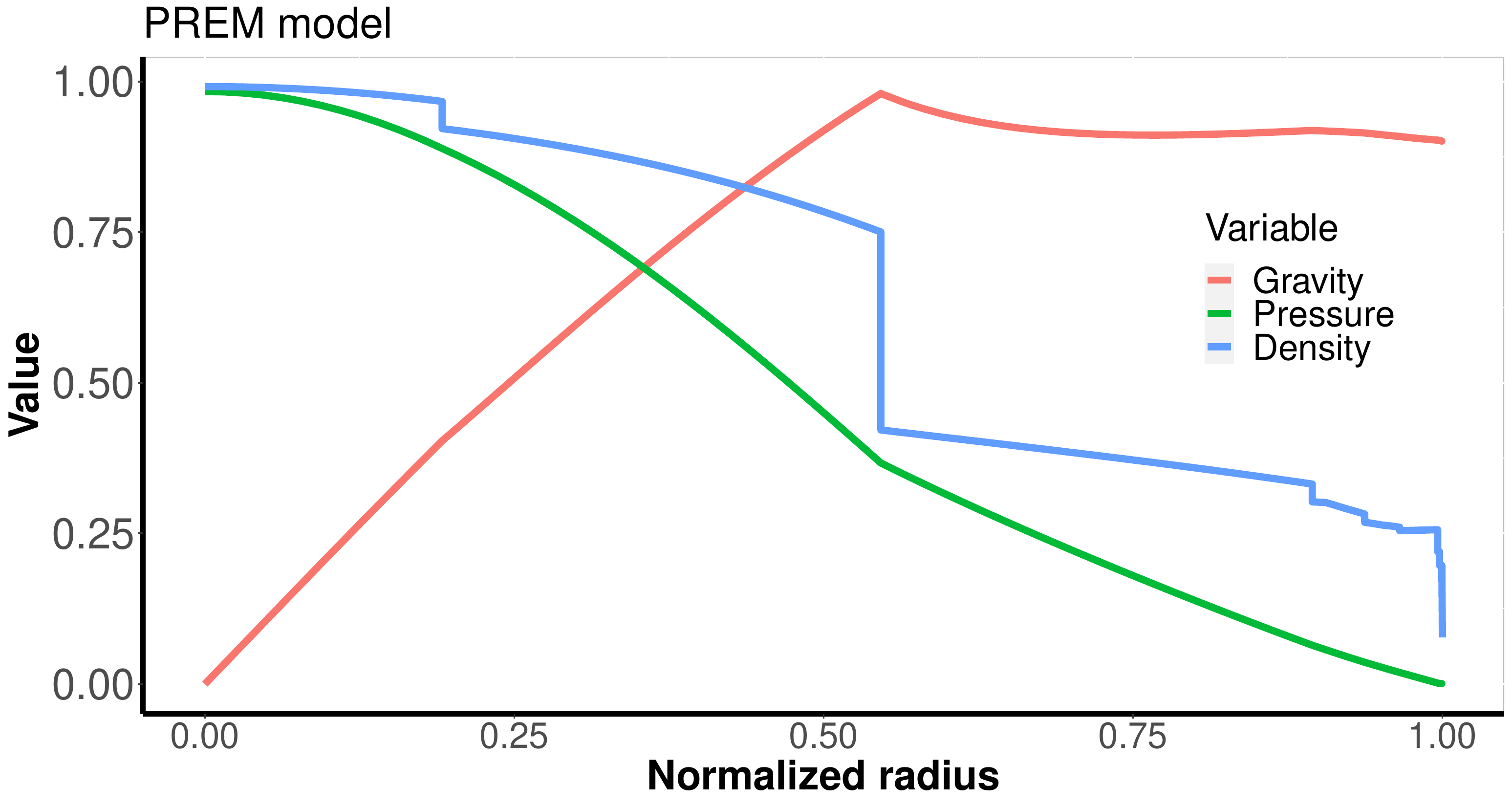}
   \caption{Density, Gravity and Pressure profiles normalized to their higher values for Earth according to PREM. $\rho_{\rm max}$ = 13.1 g cm$^{-3}$, $P_{\rm max}$ = 364 GPa, and $g_{\rm max}$ = 10.7 m s$^{-2}$.
}
              \label{prem_plot}%
    \end{figure}

According to PREM, you can distinguish five large areas in the terrestrial interior: Ocean Layer, with a mean thickness of 3 km, Upper Mantle (mean thickness of 667 km), Lower Mantle (mean thickness of 2221 km), Outer Core (mean thickness of 2258.5 km) and Inner Core (mean thickness of 1221.5 km). The most important physical characteristics of the Earth arising from PREM have been included in table \ref{Earth_prem}, while in figure \ref{prem_plot} we can see the variation with the radius of the density, gravity, and pressure provided by this model. The density profile has four different zones: upper mantle, lower mantle, outer core, and inner core, with abrupt changes in density in the upper mantle and between these zones. Concerning gravity, two zones are distinguished: the mantle where it varies very smoothly from the surface down to the CMB, and the core where it varies linearly down to zero gravity at the centre of the planet, although the slope in the outer core is slightly lower than the slope in the inner core.

   \begin{table}
      \caption[]{Earth physical characteristics according to PREM.}
         \label{Earth_prem}
         \resizebox{0.5\textwidth}{!}{%
     \begin{tabular}{c|c}
            Property & Value \\
            \hline
            Mass &	$5.974\times 10^{24}$	kg\\
            Core Mass Fraction	& 0.325 \\	 
            Mantle Mass Fraction &	0.675 \\	 
            Mean Radius &	6371 km \\
            Inner Core Radius &	1221.5 km\\
            Outer Core Radius &	3480 km\\
            Lower Mantle Radius	& 5701 km\\
            Central Planet Density	& 13.0885 g cm$^{-3}$\\
            Inner-Outer Core Boundary Density & 12.7636-12.1663 g cm$^{-3}$\\
            Core-Mantle Boundary Density &	9.9035-5.5665 g cm$^{-3}$\\
            Lower-Upper Mantle Boundary Density	& 4.3807-3.9921 g cm$^{-3}$\\
            Core Mean Density &	11.000 g cm$^{-3}$\\
            Mantle Mean Density	& 4.447 g cm$^{-3}$\\
            Mean Density	& 5.515	g cm$^{-3}$\\
            Central Planet Pressure	& 363.852 GPa\\
            Inner - Outer Core Boundary Pressure & 328.851 GPa\\
            Core - Mantle Boundary Pressure	& 135.751 GPa\\
            Lower - Upper Mantle Boundary Pressure	& 23.833 GPa\\
            Inner - Outer Core Boundary Gravity	& 4.4002 m s$^{-2}$\\
            Core - Mantle Boundary Gravity	& 10.6823 m s$^{-2}$\\
            Surface Gravity	& 9.8156 m s¯²

     \end{tabular}}
   \end{table}

\subsection{Internal structure model for dry rocky planets}
\label{PREM_model}

In general, in the literature to determine the internal structure of the planets, models based on EOS of minerals and metals obtained either theoretically or experimentally in the laboratory are used. Several previous models of solid planets have been built using pure constituents, like solid Fe-$\epsilon$ in the core and pv and ppv in the mantle, without contemplating the presence of impurities or other elements, and considering a complete differentiation of layers that do not occur on real planets \citep{Zeng08, Zeng13}. These theoretical models do not represent the Earth characteristics properly. On the one hand, the Earth's liquid outer core has a lower density than that corresponding to the solid Fe-$\epsilon$, and on the other hand, the density of the upper mantle cannot be obtained extrapolating the density of the lower mantle \citep{Zeng16}. In addition, a perfect differentiation of the different layers is not realistic.

Firstly, we are going to define a model for reproducing the internal structure of the Earth as accurately as possible and then move on to exoplanets. This accurate model for dry rocky planets is constructed under the following assumptions:

\begin{itemize}

	\item[$\bullet$] The planet has spherical symmetry.

	\item[$\bullet$] The ocean layer will not be considered because it is irrelevant from the point of view of the internal structure. On Earth, it accounts for only 0.02$\%$ of the total mass.

	\item[$\bullet$] In phase transitions of denser constituents such as iron or enstatite, the effect of temperature will be disregarded compared to the effect of pressure \citep{Valencia06}.

	\item[$\bullet$] The second-order Birch-Murnaghan EOS will be used to define the density behaviour \citep{Birch52}. This provides a precise description of how the materials are compressed inside Earth. This EOS is defined as follows:

\begin{equation}
    P=\frac{3}{2}  K_i  \Big[\Big(\frac{\rho}{\rho_i}\Big)^{7/3}- \Big(\frac{\rho}{\rho_i}\Big)^{5/3}\Big]
    \label{presion_densidad}
\end{equation}

    Where $\rho$ is the density presented by the material when subjected to a pressure $P$, $K_i$ is the isothermal compressibility, and $\rho_i$ is a reference density. This reference density is determined for each of the four layers defined for the Earth comparing the mean density reported by PREM with that calculated by our model. Following this procedure we have obtained:
    
    \begin{itemize}
        \item Solid core: $K_i$= 255 GPa, and $\rho_i$=7.848 g cm$^{-3}$
        \item Liquid core: $K_i$= 201 GPa, and $\rho_i$=7.055 g cm$^{-3}$
        \item Lower mantle: $K_i$= 206 GPa, and $\rho_i$=4.010 g cm$^{-3}$
        \item Upper mantle: $K_i$= 206 GPa, and $\rho_i$=3.329 g cm$^{-3}$
    \end{itemize}

    It is important here to highlight that for this model we have not assumed pure components for a given layer or perfect layer differentiation.
    
	\item[$\bullet$] For pressures above 12000 GPa in the core and 3500 GPa in the mantle, electron degeneration pressure dominates while the crystalline structures become less important \citep{Zeng16}. From these high pressures on, we will use the Thomas-Fermi-Dirac EOS (TFD) modified with energy correlation \citep{Salpeter67}. This EOS will provide a lower limit for the density of the material under consideration. The atomic values for the enstatite molecule in the mantle we use are $A$ = 20 y $Z$ = 10 and the corresponding for Fe at the nucleus are $A$ = 55.845 y $Z$ = 26. The use of this representation is validated by the good agreements we find when comparing with PREM, with other Mass-Radius models in the literature, and with real masses and radii, as we show in next sections.
\end{itemize}
We have compared this accurate model with PREM. The density, pressure, and gravity profiles obtained faithfully follow those defined by PREM, and the average errors in the variables are negligible. Therefore, it can be concluded that this model reproduces in a tight way the internal structure of the Earth defined by PREM, as expected. 

Our goal is to extend this model to exoplanets. In this case, we don't know whether the core is solid, liquid, or partially liquid as it is the case of the Earth. But we know that for masses larger than 2.5 M$_\oplus$ rocky planets are unable to generate a solid core \citep{Gaidos}, and for masses larger than 2 M$_\oplus$ the core remains liquid until the shutdown of its dynamo \citep{Zuluaga13, Driscoll}. Therefore, for exoplanets with masses larger than 2 M$_\oplus$, we assume that their cores are liquids with the same behaviour as the liquid part of the Earth's core. On the other hand, exoplanets with masses lower than 2 M$_\oplus$ we simplify the model assuming the core as a single layer core defined by its radius and mean density and with a pressure-density relation following that of the Earth, that is, that described at equation \ref{presion_densidad}.


   \begin{table*}
      \caption[]{Comparison between PREM and our simplified model for the Earth, treated as an exoplanet.}
         \label{Earth_exop_tab}
     \begin{tabular}{cc|ccc}
            Property & Units  & PREM	& Earth	& Difference \\
            \hline
            Total Mass &	kg &	$5.974\times 10^{24}$ &	$5.974\times 10^{24}$ &	0.002$\%$\\
           CMF	 & &	0.325 &	0.325 &	0.003$\%$\\
            MMF	 & &	0.675 &	0.675 &	0.001$\%$\\
            Core Mean Density	& g cm$^{-3}$ &	11.000 &	11.000 &	0.003$\%$\\
            Mantle Mean Density	& g cm$^{-3}$ &	4.447 &	4.448 &	0.004$\%$\\
            Planet Mean Density	& g cm$^{-3}$ &	5.515 &	5.515 &	0.002$\%$\\
            Planet Mean Pressure &	GPa &	180.7 &	181.5 &	0.365$\%$\\
            Planet Mean Gravity	& m s$^{-2}$ &	7.68 &	7.69 &	0.124$\%$
     \end{tabular}
   \end{table*}

   \begin{figure}
   \centering
    \includegraphics[width=\linewidth]{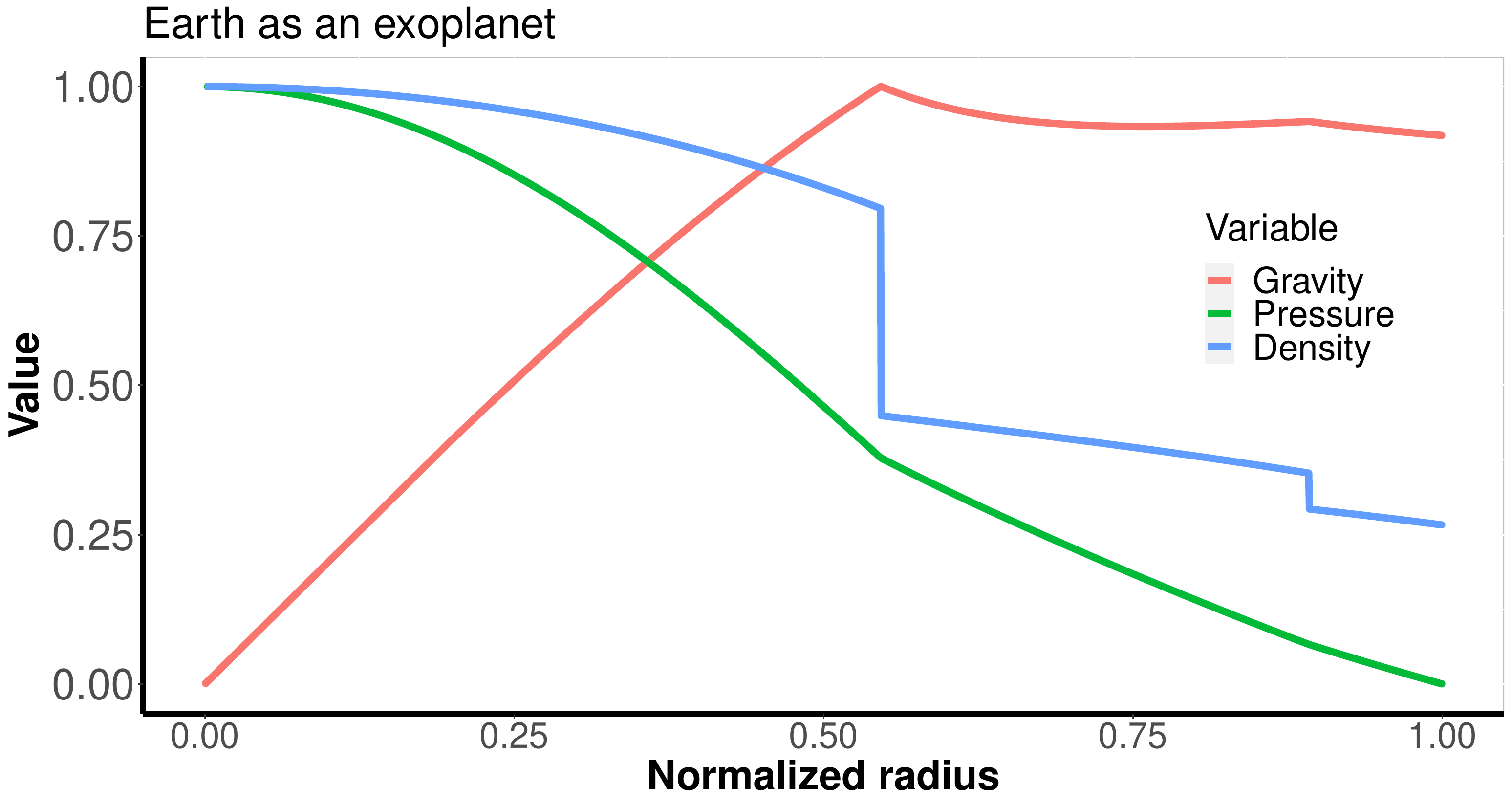}
   \caption{Density, Gravity and Pressure profiles normalized to their higher values for the Earth according our simplified model for exoplanets. $\rho_{max}$ = 12.5 g cm$^{-3}$, $P_{max}$ = 361 GPa, and $g_{max}$ = 10.7 m s$^{-2}$.
}
              \label{earth_exop}%
    \end{figure}

If we treat the Earth as an exoplanet and model its core using this simplification, when comparing with PREM we obtain that the second-order Birch-Murnaghan EOS coefficients are now $K_i$= 201 GPa, and $\rho_i$= 7.069 g cm$^{-3}$. In Fig. \ref{earth_exop} we show the density, gravity, and pressure profiles as a function of the planet radius obtained for the Earth with this simplified model and in Table \ref{Earth_exop_tab} a comparison of these results with the physical characteristics extracted from PREM. Here we can see how this simplified model can reproduce accurately the Earth's internal structure provided by PREM. Only in the core, we can see the absence of two layers in our model, substituted by a mean single layer.

\subsection{Discussion}

We can calculate the fraction of mass that represents the core relative to the total of the planet (Core Mass Fraction, CMF) like:

\begin{equation}
    {\rm CMF} = \frac{\rho_0r_0^3 }{M_p}
    \label{eq_CMF}
\end{equation}

As well as the fraction of mass of the mantle relative to the total mass of the planet (Mantle Mass Fraction, MMF):

\begin{equation}
    {\rm MMF}= \frac{\rho_1(R_p^3  - r_0^3)}{M_p}
    \label{eq_MMF}
\end{equation}

For PREM, which values are shown in table \ref{Earth_prem}, we obtain a CMF=0.325 and a MMF=0.675. On any dry rocky planet, where only the core and the mantle are significant in terms of mass, ${\rm CMF}+{\rm MMF}=1$. Therefore, from equations \ref{eq_CMF} and \ref{eq_MMF} we obtain:

\begin{equation}
    \rho_p=\rho_1+(\rho_0-\rho_1)\frac{r_0^3}{R_p^3}
    \label{eq_density}
\end{equation}

\noindent being $\rho_p$ the average normalized density of the planet.

Another key factor when establishing the internal structure of a planet is the axial moment of inertia factor that imposes the following condition on the internal distribution of mass \citep{Schubert04}:

\begin{equation}
    \frac{\rm C}{M_p R_p^2}=\frac{2}{5\rho_p}\Big[ \rho_1 +(\rho_0- \rho_1 )\frac{r_0^5}{R_p^5}\Big]
    \label{eq_int_dist_mass}
\end{equation}

   \begin{table*}
      \caption[]{Results for Dry Rocky planets of Solar System.}
         \label{Dry_Rocky_planets}
         \resizebox{1\textwidth}{!}{%
     \begin{tabular}{cc|cc|cc|cc}
     & & \multicolumn{2}{c}{Earth} & \multicolumn{2}{c}{Venus} & \multicolumn{2}{c}{Mercury}\\
     \hline
     Property & Variable & Reference & Model & Reference & Model & Reference & Model\\
     \hline
     Mass  (kg) &	$M_p$	& \multicolumn{2}{c|}{$5.974\times 10^{24}$ (1)}	& \multicolumn{2}{c|}{$4.869\times 10^{24}$ (3)}	& \multicolumn{2}{c}{$3.302\times 10^{23}$ (5)}\\ 
     Radius  (km) &	$R_p$	& \multicolumn{2}{c|}{$6371.00\pm 0.01$ (2)} &	\multicolumn{2}{c|}{$6051.8\pm 1$ (2)}	& \multicolumn{2}{c}{$2439.7\pm 1$ (2)}\\
     Average Density  (g cm$^{-3}$) &	$\rho$ &	\multicolumn{2}{c|}{5.515}	& \multicolumn{2}{c|}{5.244}	& \multicolumn{2}{c}{5.428}\\
     Core Radius (km) &	$r_0$ &	\multicolumn{2}{c|}{3480 (1)} &	3038-3292 (4) &	3258	& 1965-2035 (6) &	1996\\
     Core Mean Density (g cm$^{-3}$)	& $\rho_0$ &	11.000 &	11.000 &	10.41	& 10.43	 & 6.97-7.50 (6) &	7.26\\
     Core Mass Fraction &	CMF &	0.325 (3) &	0.325	& 0.31 (3) &	0.31	& 0.738 &	0.732 \\
     Mantle Mean Density  (g cm$^{-3}$) &	$\rho_1$ &	4.447	& 4.447 &	4.29 &	4.29	& 3.16-3.42 (6) &	3.22\\
     Mantle Mass Fraction &	MMF	& 0.675 (3) &	0.675	& 0.69 (3) &	0.69	& 0.262	& 0.268\\
     Planet  Mass  Error &&	\multicolumn{2}{c|}{0.003$\%$}	& \multicolumn{2}{c|}{0.001$\%$}	& \multicolumn{2}{c}{0.006$\%$}\\
     Average Density Error &&	\multicolumn{2}{c|}{0.001$\%$} &	\multicolumn{2}{c|}{0.001$\%$} &	\multicolumn{2}{c}{0.001$\%$}\\
     \hline
     \end{tabular}}
    \begin{tablenotes}
        \item References: (1)~\citet{Dzie81}; (2)~\citet{Zeng16}; (3)~\citet{Seidelmann07}; (4)~ \citet{Dumoulin17}; (5)~\citet{Smith12}; (6)~\citet{Rivoldini13}.
    \end{tablenotes}
   \end{table*}

   \begin{table*}
      \caption[]{Results for Low Density Dry Rocky objects of Solar System.}
         \label{LD_Dry_Rocky_planets}
         \resizebox{1\textwidth}{!}{%
     \begin{tabular}{cc|cc|cc|cc}
     & & \multicolumn{2}{c}{Mars} & \multicolumn{2}{c}{Io} & \multicolumn{2}{c}{Moon}\\
     \hline
     Property & Variable & Reference & Model & Reference & Model & Reference & Model\\
     \hline
     Mass  (kg) &	$M_p$	& \multicolumn{2}{c|}{$(6.4186\pm 0.0008) \times 10^{23}$ (1)}	& \multicolumn{2}{c|}{$(0.893 \pm 0.001) \times 10^{23}$}	& \multicolumn{2}{c}{$(0.735 \pm 0.001) \times 10^{23}$}\\ 
     Radius  (km) &	$R_p$	& \multicolumn{2}{c|}{$3389.5 \pm 0.2$ (2)} &	\multicolumn{2}{c|}{$1821.5\pm 0.5$ (2)}	& \multicolumn{2}{c}{$1737.4 \pm 1$ (1)}\\
     Average Density  (g cm$^{-3}$) &	$\rho$ &	\multicolumn{2}{c|}{$3.935 \pm 0.001$}	& \multicolumn{2}{c|}{$3.528 \pm 0.003$ (4)}	& \multicolumn{2}{c}{$3.3456 \pm 0.0004$ (7)}\\
     Core Radius (km) &	$r_0$ &	1794 $\pm$ 65 (3) & 1734 &	600-800 (6) &	780	& 340 $\pm$ 30 (8) &	478\\
     Core Mean Density (g cm$^{-3}$)	& $\rho_0$ &	5.53-7.14 &	7.0 &	6.25-7.22 (6)	& 6.29	 & 6.30-7.00 (6) &	6.25\\
     Core Mass Fraction &	CMF &	0.232-0.241 (3) &	0.238	& 0.14-0.16 (5) &	0.14	& 0.013-0.015 (8) &	0.04 \\
     Mantle Mean Density  (g cm$^{-3}$) &	$\rho_1$ &	3.44-3.62	& 3.46 &	3.18-3.31 &	3.29	& 3.32 &	3.28\\
     Mantle Mass Fraction &	MMF	& 0.759-0.768 &	0.762	& 0.84-0.86 &	0.86	& 0.985-0.987	& 0.96\\
     Planet  Mass  Error &&	\multicolumn{2}{c|}{0.003$\%$}	& \multicolumn{2}{c|}{0.001$\%$}	& \multicolumn{2}{c}{0.005$\%$}\\
     Average Density Error &&	\multicolumn{2}{c|}{0.001$\%$} &	\multicolumn{2}{c|}{0.001$\%$} &	\multicolumn{2}{c}{0.001$\%$}\\
     \hline
     \end{tabular}}
    \begin{tablenotes}
        \item References: (1)~\citet{Konopliv11}; (2)~\citet{Seidelmann07}; (3)~\citet{Rivoldini11}; (4)~\citet{Schubert04}; (5)~\citet{Kuskov01}; (6)~\citet{Kuskov16}; (7)~\citet{Williams14}; (8)~\citet{Kronrod11}.
    \end{tablenotes}
   \end{table*}

\noindent being C the axial moment of inertia. When solving the internal structure of dry rocky planets when only its mass and radius are known, we have three unknown variables ($r_0$, $\rho_0$ and $\rho_1$), and two relations among these variables (equations \ref{eq_density} and \ref{eq_int_dist_mass}). We also have The EOS of the different layers of the mantle and core in the way that, for every $r_0$ we have the mean density of the mantle and the core. We use equation \ref{eq_density} for obtaining the value of $r_0$ making null the difference between the observed density and that coming from the model. Then we use equation \ref{eq_int_dist_mass} for estimating the planetary moment of inertia factor.

\subsection{Application for other rocky planets in the solar system}

As the first test of our model, we have verified whether it is capable of reproducing the internal structure of the other rocky planets in the solar system. To do this, the mass, radius, and radius of the nucleus of Mercury, Venus, and Mars have been introduced as input data in our model. The results obtained can be seen in Table \ref{Dry_Rocky_planets}, where we can see that the internal structure of Mercury and Venus is reproduced correctly with minimal errors.

   \begin{figure}
   \centering
    \includegraphics[width=\linewidth]{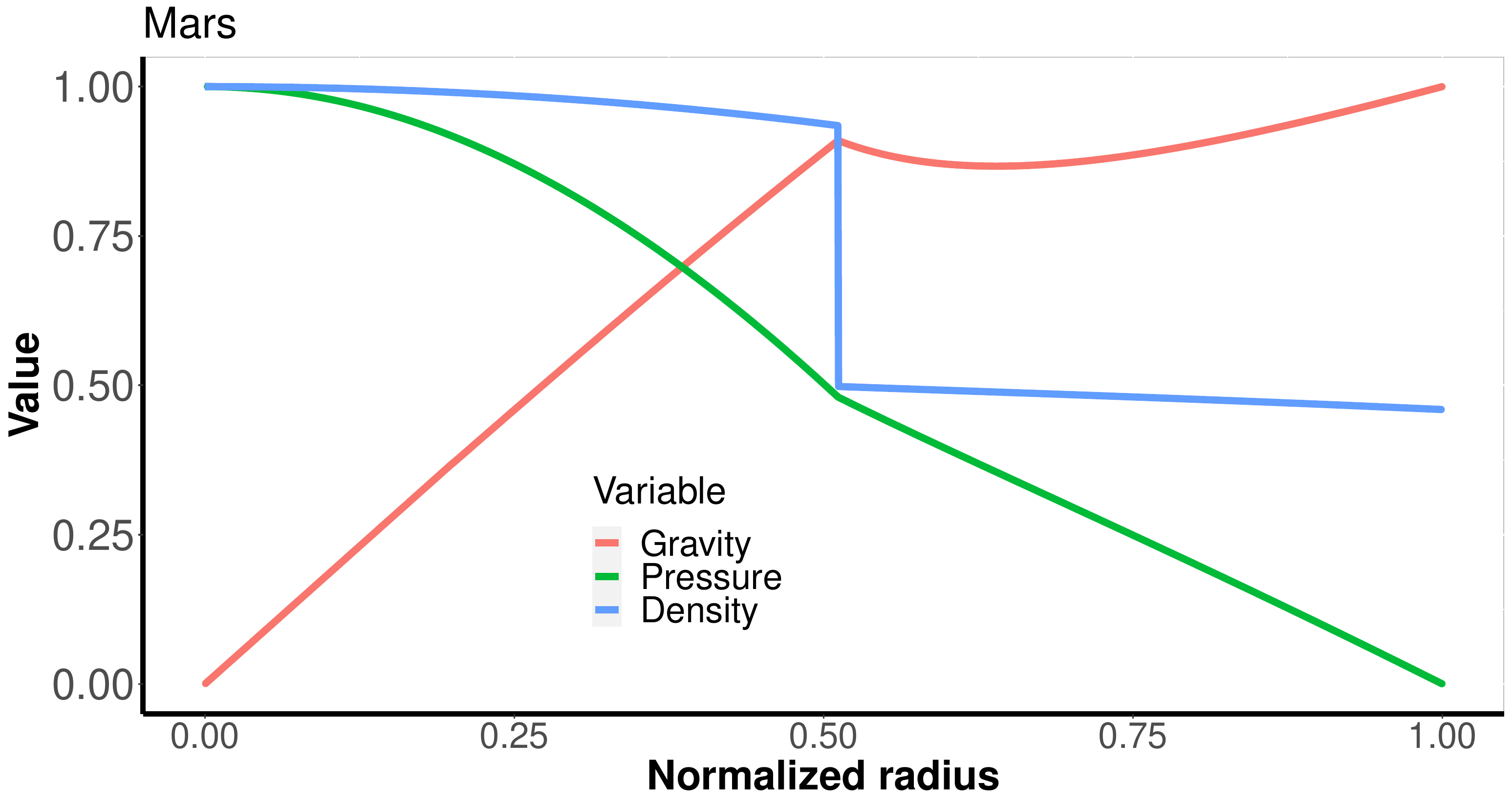}
   \caption{Density, pressure, and gravity normalized to their higher values of Mars as a function of the radius obtained with our model. $\rho_{max}$ = 7.28 g cm$^{-3}$, $P_{max}$ = 40.7 GPa, and $g_{max}$ = 3.7 m s$^{-2}$.}
              \label{mars_plot}%
    \end{figure}

However, it is not the same with Mars where errors are greater. There are two very different observational data between Mars and Earth. Mars has an average density of 29$\%$ lower than Earth while the axial moment of inertia factor is 9$\%$ higher. Both data induce thinking of a difference in density between the core and the mantle of Mars is much smaller than for the Earth. This might be because during the formation of Mars, which took place farther from the Sun compared to the Earth, lighter materials have been attached to its core, like S that can reach up to 16$\pm$2$\%$ of the mass of the Mars core \citep{Rivoldini11}. On the other hand, its larger moment of inertia factor suggests a lower differentiation between mantle and core densities in Mars as compared to the Earth.

To obtain the internal structure of Mars we have used the same model modifying only the parameters making null the difference with the real mean densities at the core and the mantle:

\begin{itemize}
    \item Low density core: $K_i$= 201 GPa, and $\rho_i$=6.263 g cm$^{-3}$
    \item Low density mantle: $K_i$= 206 GPa, and $\rho_i$=3.343 g $cm^{-3}$
\end{itemize}

This modified model of low-density exoplanets has been applied to Mars, Io, and The Moon. The results obtained can be seen in the table \ref{LD_Dry_Rocky_planets}, where it can be verified that the modified model represents correctly the internal structure of these objects. The density, pressure, and gravity obtained for Mars with the modified model as a function of the radius can be seen in Figure \ref{mars_plot}. The boundary between what is a low-density exoplanet or not is not well defined yet. Mars has a density a 71$\%$ of the Earth and our low-density exoplanet model works for Mars. Therefore, this boundary must be between 71$\%$ and 99$\%$ the Earth density. We have used a value of 80$\%$ for this boundary, but it must be fine-tuned with additional observations.

\section{Internal structure of water-rich rocky planets}

Water-rich rocky planets, also called ice planets, are thought to be formed beyond the snow-line and mostly contain water and silicates. Some of these ice planets may migrate into the inner area of the stellar system by interaction with the stellar disk or another planet \citep{Kuchner03}. If the migration eventually produces an orbit within its stellar HZ, we would have an ocean planet. This does not mean that the entire surface of the planet is covered by water since polar caps could remain covered by ice. The size of the polar ice caps will depend on the effective flux coming from the star. To study the internal structure of the ice planets we will use the objects of this type available closest to the HZ of Sun, that is, the Galilean´s moons of Jupiter.

\subsection{Structure Model}
\label{model_wr}

In general, a water-rich rocky planet consists of three layers (see Fig. \ref{Estruct_int}). To the structure of a dry rocky planet, we add an ice cap that can also contain an inner ocean. Taking into account that these objects have formed beyond the snow-line and are low-density objects, we have assumed that the internal structure of these planets is more similar to that of Mars. To this model, we have patched a new layer composed of ice. The EOS representing light elements like ice cannot ignore the effect of temperature. Therefore, we have implicitly included a well-known pressure-temperature relation such as the water melting curve \citep{Zeng13}. In our model to define the internal structure of ice objects we have used the following elements:

\begin{itemize}
    \item The mantle and core density have a behaviour similar to those of Mars.
    \item We have used the water phase diagram from \citet{Choukroun10}. This diagram defines the pressures at which ice changes its crystallization system.
    \item For ice Ih, ice III, ice V, and ice VI we have used the EOS defined by \cite{Gagnon90}.
    \item For ice VII we have applied the EOS proposed by \cite{Frank04}.
    \item For ice X we have used the EOS obtained by \cite{French09}.
    \item We have used the EOS proposed by \cite{Vance18} for salted water with a 10$\%$ wt of MgSiO$_4$. This is done because Galileo spacecraft has detected the very likely existence of saltwater oceans in Jupiter's ice moons.
\end{itemize}

\subsection{Discussion}

Assuming a water-rich rocky planet with a normalized mass $M_p$ and a normalized radius $R_p$, the CMF is still defined by the equation \ref{eq_CMF}. The MMF now follows:

\begin{equation}
    {\rm MMF}=\frac{\rho_1(r_1^3  - r_0^3)}{M_p}
    \label{eq_MMF2}
\end{equation}

\noindent where $r_1$ is the normalized radius of the mantle, and $\rho_1$ its normalized mean density.

The mass fraction of the ice layer (Ice Mass Fraction, IMF) can be defined as:

\begin{equation}
    {\rm IMF}=\frac{\rho_2(R_p^3  - r_1^3)}{M_p}
    \label{eq_IMF}
\end{equation}

\noindent where $\rho_2$ its normalized mean density of this layer.

Therefore, in this case these coefficients must verify that $CMF+MMF+IMF=1$. Including equations \ref{eq_CMF}, \ref{eq_MMF2}, and \ref{eq_IMF} in this expression we obtain:

\begin{equation}
    \rho_p=\rho_2+(\rho_0-\rho_1)\frac{r_0^3}{R_p^3}+(\rho_1-\rho_2)\frac{r_1^3}{R_p^3}
    \label{eq_density2}
\end{equation}

\noindent being $\rho_p$ the mean normalized density of the planet. Finally, the axial moment of inertia factor now it is described as \citep{Schubert04}:

\begin{equation}
    \frac{\rm C}{M_p R_p^2}=\frac{2}{5\rho_p}\Big[ \rho_2 +(\rho_0- \rho_1 )\frac{r_0^5}{R_p^5}+(\rho_1- \rho_2)\frac{r_1^5}{R_p^5}\Big]
    \label{eq_int_dist_mass2}
\end{equation}

To solve the internal structure of water-rich rocky planets, we have in this case equations \ref{eq_density2} and \ref{eq_int_dist_mass2} and five unknown variables ($r_0$, $r_1$, $\rho_0$, $\rho_1$ and $\rho_2$). Consequently, additional data are needed. In the case of Jupiter's ice moons their moment of inertia factor, sub-surface conductivity of the planet, gravitational constants like C$_{22}$, as well as the EOS of the core, mantle, and ice layer constituents, help us close the problem.

When analyzing water-rich rocky exoplanets, some of these additional variables, such as the moment of inertia factor,  are unknown. In this case, for each mass and radius value, the theoretical models provide infinite solutions for the internal structure (degenerate solution) \citep[]{Suissa18, Zeng13}. To solve this degeneracy, for this type of exoplanet it is only possible to extrapolate the internal structure of some known ice object from the solar system.

\subsection{Ganymede's internal structure}

Ganymede is the biggest satellite in the solar system with a radius of 2632.3 km, greater even than Mercury \citep{Seidelmann07}. Data from the Galileo spacecraft have provided information about its internal structure. Its moment of inertia factor (0.3115) is the lowest of all known solid objects in the solar system and reveals a complete differentiation of the planet into three layers: the first layer of ice and water, then a rocky mantle, and finally a metal core \citep{Schubert04}. Its low average density (1.942 $g\,cm^{-3}$) suggests an important content in water and ice so a high IMF value is expected. Recent internal structure models of this object suggest a thickness of the ice and water layer around one-third of the planet's radius, i.e., between 800-900 km for \citep{Kuskov16} and between 876-910 km for \citep{Vance18}. Galileo spacecraft magnetometer data reveals a relatively strong dipolar magnetic field (0.002 $\mathcal{M}_\oplus$) that must be generated by the action of a dynamo in a liquid or partially liquid core \citep{Schubert04}. \citep{Kuskov16}, has calculated, with 10 GPa / 2000 K in the centre of the satellite, a core size between 600-800 km and an average density between 6.6 - 7.4 cm$^{-3}$ that would imply a CMF of between 0.04 and 0.10.

A response induced to Jupiter's magnetic field has also been detected that requires the existence of a conductive layer compatible with a saltwater ocean \citep{Kivelson02}. The Galileo spacecraft has also provided a relatively low value for the gravitational constant C$_{22}$ measuring Ganymede's tidal response that may mean that tidal warming is negligible \citep{Bland15}. This may show that the salty ocean would not have to be very close to the surface. 

Introducing in our model all the above observational data and iterating until a null mean density difference with the observed value is reached, we obtain:
\begin{itemize}
    \item $r_0$= 770 km
    \item $D_{ice} = R_p- r_1 = 901$ km
    \item $D_{ocean}$= 130 km
\end{itemize}

\begin{figure}
   \centering
    \includegraphics[width=\linewidth]{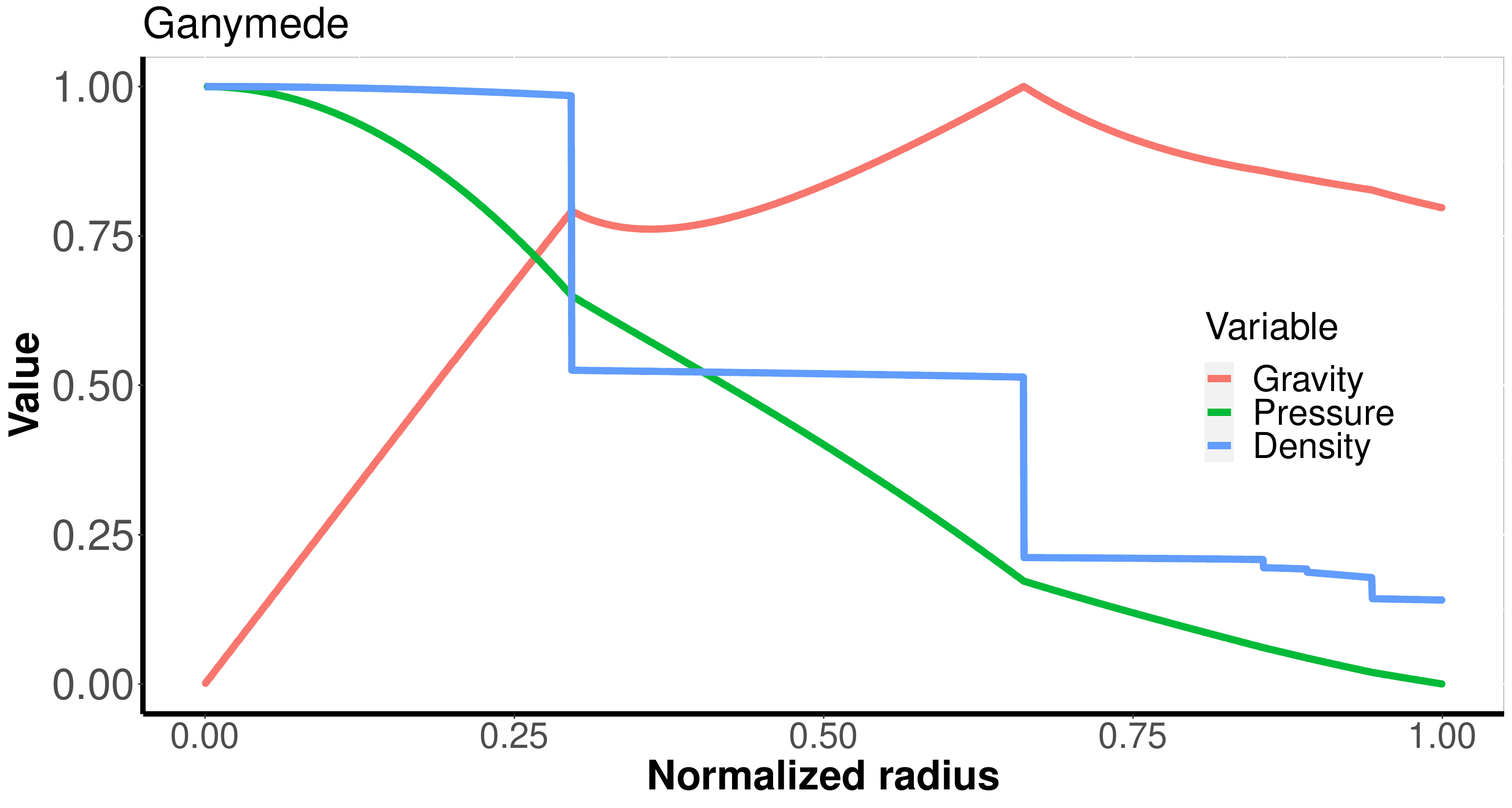}
   \caption{Density, pressure, and gravity normalized to their higher values of Ganymede as a function of the radius obtained with our model. $\rho_{max}$ = 6.6 g cm$^{-3}$, $P_{max}$ = 10.3 GPa, and $g_{max}$ = 1.8 m s$^{-2}$.
}
              \label{ganimede_plot}%
    \end{figure}

The complete results can be seen in Table \ref{ice_satellite}, and the density profile obtained for Ganymede in Figure \ref{ganimede_plot}.

   \begin{table*}
      \caption[]{Results for Ice satellites of Solar System.}
         \label{ice_satellite}
         \resizebox{1\textwidth}{!}{%
     \begin{tabular}{cc|cc|cc|cc}
     & & \multicolumn{2}{c}{Ganymede} & \multicolumn{2}{c}{Europa} & \multicolumn{2}{c}{Callisto}\\
     \hline
     Property & Variable & Reference & Model & Reference & Model & Reference & Model\\
     \hline
     Mass  (kg) &	$M_p$	& \multicolumn{2}{c|}{$1.483 \times 10^{23}$}	& \multicolumn{2}{c|}{$4.772 \times 10^{22}$}	& \multicolumn{2}{c}{$1.075 \times 10^{23}$}\\ 
     Radius  (km) &	$R_p$	& \multicolumn{2}{c|}{2632.3 (1)} &	\multicolumn{2}{c|}{1562.1 (1)}	& \multicolumn{2}{c}{2409.3 (1)}\\
     Average Density  (g cm$^{-3}$) &	$\rho$ &	\multicolumn{2}{c|}{$1.942  \pm 0.005$ (2)}	& \multicolumn{2}{c|}{$2.989  \pm 0.005$ (2)}	& \multicolumn{2}{c}{$1.834  \pm 0.003$ (2)}\\
     Normalized Moment of Inertia factor&	$\frac{\rm C}{M_p R_p^2}$	& \multicolumn{2}{c|}{$0.3115  \pm 0.003$ (2)}	& \multicolumn{2}{c|}{$0.3460  \pm 0.005$ (2)} & 	\multicolumn{2}{c}{0.3200 (6)}\\
     \hline
     \multirow{3}{2.4cm}{\centering Core Radius (km)} &	\multirow{3}{1.5cm}{\centering $r_0$} &	600-800 (3) & \multirow{3}{1.5cm}{\centering 770} & 469-662 (2) &	\multirow{3}{1.4cm}{\centering 621}	& \multirow{3}{1.5cm}{\centering 568-592(2)} &	\multirow{3}{1.4cm}{\centering 665}\\
     &&\multirow{2}{1.5cm}{\centering 501-602(4)}&&478 (4)&&&\\
     &&&&470-640 (5)&&&\\
     \multirow{2}{2.4cm}{\centering Ice-water Thickness (km)} &	\multirow{2}{1.5cm}{\centering $R_p-r_1$} &	800-900 (3) & \multirow{2}{1.4cm}{\centering 901} &	136 (4) &	\multirow{2}{1.4cm}{\centering 121}	& \multirow{2}{1.5cm}{\centering --} &	\multirow{2}{1.4cm}{\centering 880}\\
     &&876-910 (4)&&125-140 (5)&&&\\
     \multirow{2}{2.4cm}{\centering Ocean Thickness  (km)} &	\multirow{2}{1.4cm}{\centering D$_{ocean}$} &	\multirow{2}{1.5cm}{\centering 24-287 (4)}& \multirow{2}{1.4cm}{\centering 130} &	129-134 (2) &	\multirow{2}{1.4cm}{\centering 119}	& \multirow{2}{1.5cm}{\centering 20-132 (2)} &	\multirow{2}{1.4cm}{\centering 120}\\
     &&&&133-136 (4)&&&\\
    \hline
     Core Mean Density (g cm$^{-3}$) &	$\rho_0$ &	6.6-7.4 (3)& 6.59 &	6.3-7.0 (3) &	6.19	& -- &	6.22\\
     Core Mass Fraction &	CMF &	0.04-0.10 &	0.085	& 0.05-0.13 (3) &	0.13	& 0.05-0.06 (2) &	0.07 \\
    \hline
     Mantle Mean Density  (g cm$^{-3}$) &	$\rho_1$ &	3.51 (4) & 3.44 &	3.425 (4) &	3.26	& 3.53  (4)&	3.27\\
     Mantle Mass Fraction &	MMF	& 0.47 &	0.46	& 0.84 &	0.79	& --	& 0.42\\
    \hline
     Ice Mean Density  (g cm$^{-3}$) &	$\rho_2$ &	1.22 & 1.23 &	1.12 &	1.14	& -- &	1.26\\
     Ice Mass Fraction &	IMF	& 0.45 &	0.455	& 0.09 &	0.08	& --	& 0.51\\
         \hline
     Planet  Mass  Error &&	\multicolumn{2}{c|}{0.014$\%$}	& \multicolumn{2}{c|}{0.026$\%$}	& \multicolumn{2}{c}{0.008$\%$}\\
     Average Density Error &&	\multicolumn{2}{c|}{0.001$\%$} &	\multicolumn{2}{c|}{0.001$\%$} &	\multicolumn{2}{c}{0.001$\%$}\\
     Moment of Inertia Error &&	\multicolumn{2}{c|}{0.032$\%$} &	\multicolumn{2}{c|}{0.003$\%$} &	\multicolumn{2}{c}{0.004$\%$}\\
     \hline
     \end{tabular}}
    \begin{tablenotes}
        \item References: (1)~\citet{Seidelmann07}; (2)~\citet{Schubert04}; (3)~\citet{Kuskov16}; (4)~\citet{Vance18}; (5)~\citet{Kuskov05}; (6) \citet{Gao13}.
    \end{tablenotes}
   \end{table*}

\subsection{Europa and Callisto internal structure}

Europa is a much smaller ice object than Ganymede, with a radius of 1562.1 km \citep{Seidelmann07}, but with a much higher mean density 2.989 g cm$^3$ \citep{Schubert04} suggesting a not too deep ice layer. The observed moment of inertia factor (0.346) requires a concentration of mass in the centre of the object and a differentiation in layers with a metal core of Fe, a mantle of silicates, and an outer ice and water layer \citep{Schubert04}. The Galileo spacecraft has detected electrical conductivity within the first 200 km deep in Europa indicating the presence of an ocean of saltwater \citep{Schubert04}. In addition, it observed a value for the gravitational constant of C$_{22}$, which is compatible with an ocean of thickness around 130 km, suggesting significant tidal warming that may have turned much of the Ih ice cap into a liquid ocean.

\citep{Kuskov05} propose a core size for Europa between 470 and 640 km, whereas \citep{Kuskov16} estimate a range for the core densities between 6.3 and 7 g cm$^{-3}$. For a concentration of S in the range of 3.5 and 10 $\%$ wt in its liquid core, and a pressure of 5 GPa and a temperature of 2000 K at its centre, the CMF ranges between 0.05 and 0.13.  

In Table \ref{ice_satellite} we show the results for Europa obtained introducing in our model all this observational data.

   \begin{figure}
   \centering
    \includegraphics[width=\linewidth]{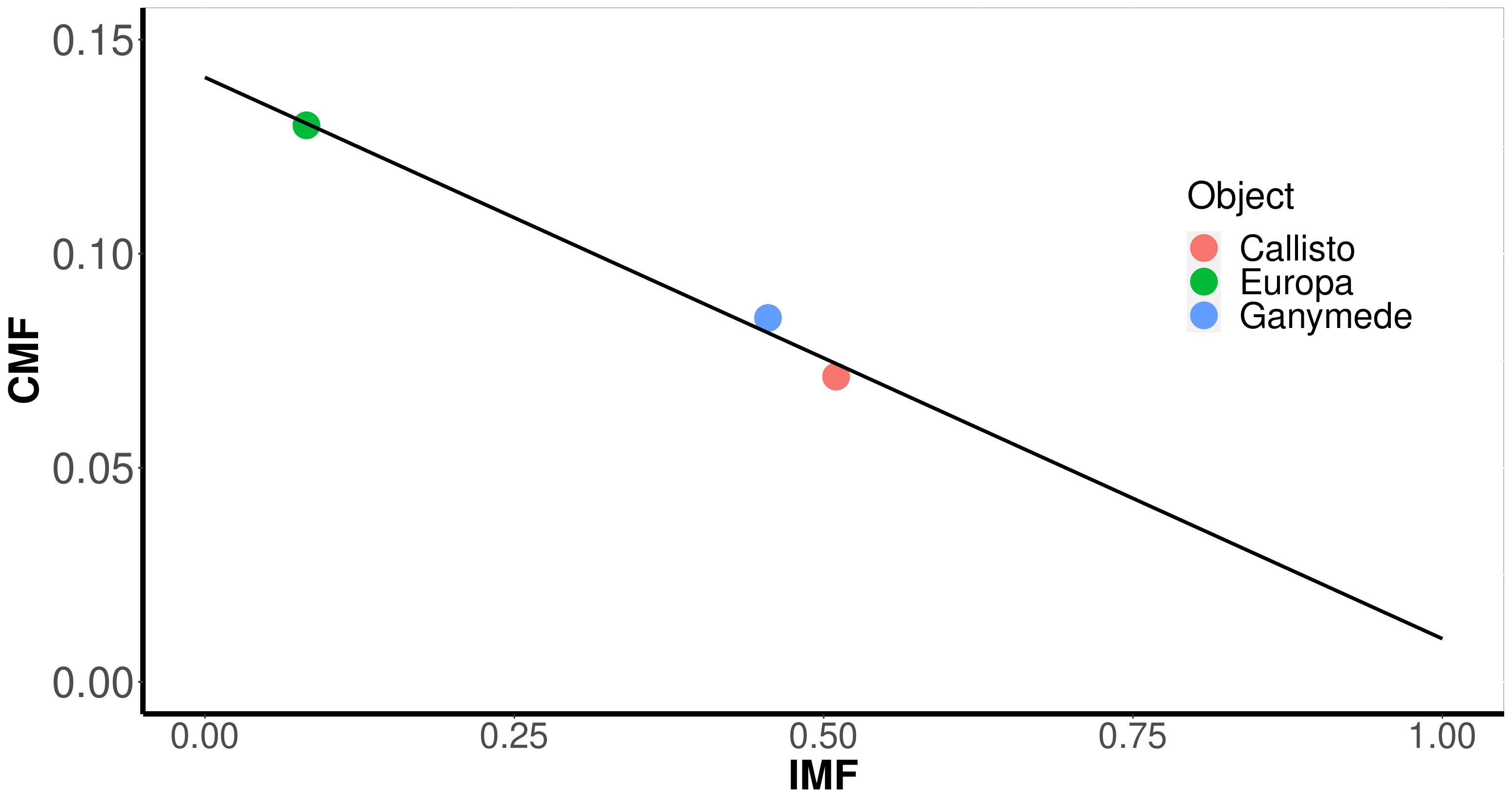}
   \caption{IMF vs. CMF for the three ice moons of Jupiter analysed: Ganymede, Europa, and Callisto. the black line represent the linear regression of these three points, to guide the eye.
}
              \label{IMF-CMF}%
    \end{figure}

Callisto is a satellite similar in size to Ganymede (radius of 2409.3 km) \citep{Seidelmann07}, but with a density even lower than it (1.8344 g cm$^3$) \citep{Schubert04} suggesting an important thickness of the ice layer. Since the Galileo spacecraft only performed equatorial flights over Callisto, the moment of inertia factor has been obtained using the \citep{Gao13} correction. Its value of 0.32 suggests a mass distribution similar to Ganymede, that is, well-differentiated layers with a large ice layer, a rocky mantle, and a metallic core. Callisto can have also an inner ocean \citep{Zimmer00}.

Applying our model, with the previous observational data, the results included in Table \ref{ice_satellite} have been obtained. Therefore, our model for ice planets produces consistent results with the data observed for these two objects.

\subsection{IMF - CMF relation}

In the light of the results we have obtained for the ice moons of Jupiter: Ganymede, Europa, and Callisto, we have found a possible relationship between their IMF and CMF. When IMF increases, CMF decreases. This can be seen in Fig. \ref{IMF-CMF}. These three observational points lie in a line. We have also plotted the regression line of these three points to guide the eye. Of course, with only three points it is impossible to assess a physical event, but we can use this property for breaking the degeneracy of our model for these objects. If we impose this condition to the model of water-rich exoplanets, for a given $M_p$ and $R_p$ we would have a unique solution compatible with the internal structure of these moons.

\section{Mass-radius grid}

Extrapolating the three internal structure models described above for Earth, Mars, and Ganymede we have constructed a new grid of exoplanets characteristics, shown in table \ref{Grid}. Thus, entering the mass and radius of an exoplanet in the grid can be quickly identified if it is a dry or water-rich rocky exoplanet, as well as its CMF or its IMF. For low-density dry rocky planets, the internal structure model of Mars has been used only when the mean density of the planet is less than 0.8 times the mean density of the Earth.

We have assumed that for a given value of a planet mass, it is a dry rocky planet when its radius is less than or equal to the radius corresponding to a 100$\%$ of rock (CMF = 0). If the radius of the planet is greater than this value, its internal structure cannot be explained with only two layers and there must be additional lighter components, such as ice. 

The grid has been made for planetary masses from 0.1 up to 48 M$_\oplus$. When extrapolating the model for dry rocky planets with masses larger than 32 M$_\oplus$, pressures are so intense that the most widely used EOS is the modified TFD, i.e. electron degeneration pressure dominates. Nevertheless, this EOS must be taken with caution since it can introduce elevated errors. Something similar happens for water-rich planets and masses larger than 48 M$_\oplus$, where the electron degeneration pressure dominates, especially for IMF<0.5. When applying our model we must take into account that all the reference objects have masses lower than the Earth. That means that for exoplanets with masses larger than 1M$_\oplus$ it extrapolates somehow the Earth's internal structure. When mass increases, the thermal conductivity, and energy flow are the most impacting factors. In this case, we use the most updated models for describing them as a function of the mass. We think this is the best approximation available nowadays, although we recognise that the extrapolation of PREM models to objects more massive than the Earth induces significant uncertainty.

\section{Magnetic properties of exoplanets}

The main goal of our study is to estimate the magnetic properties of exoplanets using the most common known characteristics such as exoplanetary mass, radius and orbital period. In this section, we describe how we obtain these estimations assuming we have a model of its internal structure as shown in the previous sections.

\subsection{Magnetic moment}
\label{Regime_and_Magnetic_moment}

Most planetary dynamo is thought to be maintained by thermal and compositional convection mechanisms in electrically conductive fluids inside planets (OC06). The scaling laws of OC06 make it possible to determine the regime (dipolar/multipolar) and magnetic moment of a planetary dynamo once the Rayleigh number is known. From this value, applying these laws, and the Rossby number, the Reynolds magnetic number, and the dipole field Lorentz number is it possible to determine the magnetic behaviour of the planet.

The Rayleigh number ($Ra_{\rm Q}$) can be defined as:

\begin{equation}
    Ra_{\rm Q}=  \frac{r^* F}{D^{2} \Omega^{3}}
\end{equation}

\noindent where $r^*$ is the ratio between the core radii ($r^*=r_0/r_i$), $D$ the core convective zone thickness ({$D=r_0-r_i$}), $\Omega$ the planet's rotational frequency and $F$ the mean buoyancy convective flux that can be obtained following:

\begin{equation}
    F=\frac{\alpha g q_{conv}}{\rho C_p}
    \label{F}
\end{equation}

\noindent being $\alpha$, $g$, $\rho$ and $C_p$ the thermal expansion coefficient, gravity, density, and the specific heat capacity at constant pressure at the planet core. While $q_{conv}$ is the convective heat flux generated by the core \citep[OC06 and][]{Driscoll}.

Another fundamental parameter to define the magnetic regime of the planet is the Rossby number:

\begin{equation}
    Ro=  \frac{u}{\Omega D}
    \label{rossby}
\end{equation}

\noindent being $u$ the speed of the fluid that generates the planetary dynamo. Similarly, the local Rossby number is defined based on the order of the spherical harmonics when spherical symmetry is assumed for the planet, that is: 

\begin{equation}
    Ro_{\ell}=  \frac{\ell_u}{\pi}Ro
\end{equation}

\noindent where $\ell_u$ is the spherical harmonic degree $\ell$ of the velocity vector $\vec{u}$.Two completely differentiated regimes, dipolar and multipolar, have been found in the action of planetary dynamos. On the one hand, in the dipolar regime, the planet's magnetic field is strong and is dominated by the dipolar component. It corresponds to low values of the local Rossby number. On other hand, the multipolar regime corresponds to high values of the local Rossby number. In this case, the magnetic field is dominated by multipoles and its value decreases drastically.

In the case of base-heated dynamos, as is the case on Earth, the transition from the dipolar regime to the multipolar regime occurs at a very narrow interval around a value of the local Rossby number of 0.12 (OC06). Values larger than 0.12 indicate a multipolar regime whereas lower values involve a dipolar regime. When the value of the local Rossby number is in a range close to and below 0.12, the planet is in an area with a dipole magnetic field with reversible polarity.

Using the scaling laws described at OC06, the local Rossby number can be estimated using the following equation:

\begin{equation}
    Ro_{\ell}=\frac{0.58}{\nu^{1/3}}\Big(\frac{\lambda_m}{\lambda_t}\Big)^{1/5}Ra_{\rm Q}^{2/5}\Omega^{1/3}D^{2/3}
\end{equation}

\noindent where $\nu$ is the kinematic viscosity, $\lambda_m$, and $\lambda_t$ are respectively the diffusivities magnetic and thermal of the fluid. Therefore, depending on the value of $Ro_{\ell}$ with respect to the critical value of 0.12 we can estimate the magnetic regime of the exoplanet. This value depends mainly, therefore, on the Rayleigh number, $\Omega$, and $D$, and weakly on the core thermodynamic properties.

The scaling laws of OC06 also allow us to know the global Rossby number that for dynamo heated by the base can be calculated by the following equation:

\begin{equation}
 Ro=  \beta Ra_{\rm Q}^{2/5}
\end{equation}

\noindent being $\beta\approx$ 0.85 (OC06). In this way, by applying equation \ref{rossby} the velocity of the fluid generated by the dynamo can be determined.

The third parameter that defines a planetary dynamo is the magnetic Reynolds number, which is determined  as:

\begin{equation}
    R_m=\frac{uD}{\lambda_m}
\end{equation}

For a planet to have a dynamo, it is necessary to have a layer of electric conductor fluid, and a Reynolds magnetic number ($R_m$) larger than 40 \citep{Gaidos}. Below that critical value, the magnetic moment is null. In this study, a prior check of the magnetic Reynolds number will always be carried out to ensure that the action of the dynamo has begun.

The fourth parameter that defines planetary dynamo is the dipole field Lorentz number: 

\begin{equation}
    Lo_{dip}=\sqrt{\frac{2\mu_0}{\rho}}\frac{\mathcal{M}}{4\pi r_0^3\Omega D}
    \label{lorentz2}
\end{equation}

\noindent where $\mu_0$ if the magnetic vacuum permeability, and $\mathcal{M}$ the magnetic moment. Scaling laws of OC06 make it easy to determine the Lorentz number from the Rayleigh number, using the following  relation:

\begin{equation}
    Lo_{dip}=\gamma_d Ra_{\rm Q}^{1/3}
    \label{lorentz}
\end{equation}

\noindent where $\gamma_d$ is the dipolar saturation constant that, for fast rotators dynamos it is assumed to have a value of 0.2 \citep{Driscoll}. Relations \ref{lorentz2} and \ref{lorentz} allows obtaining an estimation of the magnetic moment as

\begin{equation}
    \mathcal{M}=\frac{4\pi\gamma_d}{\sqrt{\mu_0}}\rho_0^{1/2}r_0^3D^{1/3}F^{1/3}
    \label{MM_zona2}
\end{equation}

If we analyse how the dipole moment varies with the Rayleigh number (see Fig. 4 of OC06), it can be observed that there are four different zones. In the first zone, for low values of $Ra_{\rm Q}$, the magnetic moment is null due either to the lack of convection or because the action of the dynamo has not begun. This zone corresponds to Reynolds magnetic number values less than 40. Once the critical value of the Reynolds number is exceeded, the dipole magnetic moment dominates. After an initial and rapid rise, a zone occurs with a linear increment of the magnetic moment relative to the Rayleigh number. In this zone, the magnetic moment can be determined by equation \ref{MM_zona2}. It is important to note that, in this zone, it does not present an explicit dependence on the angular velocity of the planet. If we continue increasing the Rayleigh number, the magnetic moment begins to be of reversible polarity, in a relatively narrow area and before reaching the critical local Rossby number of 0.12. Here is where a maximum value of the magnetic moment ($\mathcal{M}^+$) is produced, and it can be calculated imposing the condition $Ro_\ell$ = 0.12, resulting:

\begin{equation}
    \mathcal{M}^+=\frac{3\gamma_d}{\sqrt{\mu_0}}{\nu^{2/9}}\Big(\frac{\lambda_t}{\lambda_m}\Big)^{2/15}\rho_0^{1/2}r_0^3D^{5/9}\Omega^{7/9}
    \label{MM_zona3}
\end{equation}

At the fourth and last zone, with values of the Rayleigh number above the critical value of 0.12, the magnetic moment suffers a fast decrease changing from dipolar to multipolar. In the cases studied by OC06 this reduction of the magnetic moment is very important, being of the order of 0.05 times the maximum value of the magnetic moment. Some other authors work with a dipole moment reduction coefficient of 0.15 \citep{Griessmeier09}. In this work, it has been verified that the reducing coefficient in the solar system is about 0.06 that comes to ratify the value determined by OC06. Therefore, the magnetic moment in this zone can be determined in relation to the maximum dipole moment as:

\begin{equation}
    \mathcal {M}^-\approx 0.05 \mathcal{M}^+
\end{equation}

\subsection{Application for exoplanets}

If we want to extrapolate this model to rocky exoplanets, several previous considerations must be made. First of all, we know that for masses greater than 2 M$_\oplus$ the core remains completely liquid, at least, until the shutdown of its dynamo \citep{Zuluaga13, Driscoll}. In this case, the thickness of the convective zone shall be considered to be equal to the radius of the core ($D$ = $r_0$).

One of the main ingredients of these equations is the exoplanet angular velocity. Unfortunately, this variable cannot be observed yet, and we need other considerations for estimating its value. More common observational data are the orbital period and eccentricity, and the hosting star mass. With these values we can estimate whether the planet is tidally coupled to the star \citep{Griessmeier09}, using it for estimating this angular velocity as follows:

\begin{itemize}
    \item If the planet is tidally coupled, we can estimate the most probable spin-orbit resonance following \citet{Dobrovolskis07}, and then, its angular velocity.
    \item If the planet is not tidally coupled, we assume free rotation leading to a dipolar magnetic moment, like of all the objects of the solar system with this characteristic. In this case we use equation \ref{MM_zona2} for estimating $\mathcal{M}$.
\end{itemize}

Other important ingredients for estimating $\mathcal{M}$ for exoplanet are the thermal and magnetic diffusivities at the exoplanetary core, $\lambda_t$ and $\lambda_m$ respectively, and their evolution with the pressure. For our model, we have initially obtained the values for the thermal and electric conductivities ($\kappa$ and $\sigma$ respectively) in the case of the Earth (see Table \ref{eath_term}), and then we have estimated their variation with pressure using \citet{Pozzo}. A more detailed description of how we do this can be found in the Appendix. Finally, we obtain the thermal and magnetic diffusivities as

\begin{eqnarray}
\lambda_t=\frac{\kappa}{\rho C_p}\\
\lambda_m=\frac{1}{\mu_0\sigma}
\end{eqnarray}

   \begin{table*}
      \caption[]{Thermodynamic parameters of the Earth.}
         \label{eath_term}
     \begin{tabular}{c|c|c|c}
            Zone	& Parameter	& Value	& Reference \\
            \hline
            \multirow{8}{1.4cm}{\centering Core} &	Thermal expansivity ($\alpha$) & $1.3\cdot 10^{-5}$ K$^{-1}$ & A\\
           	& Specific heat capacity ($C_p$) & 850 J kg$^{-1}$ K$^{-1}$ & A B C\\
            & Thermal conductivity ($\kappa$) & 60 W m$^{-1}$ K$^{-1}$ & F\\
            & Thermal diffusivity ($\lambda_t$) & $1\cdot 10^{-5}$ m$^2$ s$^{-1}$ & F\\
            & Electric conductivity ($\sigma$) & $1.36\cdot 10^6$ s m$^{-1}$ & D\\\
            & Magnetic diffusivity ($\lambda_m$) & 1.32 s$^{-1}$ m$^2$ & D\\
            & Adiabatic factor ($\epsilon_{\rm ad}$) & 0.7 & C\\
            & $\theta_c$ coefficient  & 0.4 & C\\
            \hline
            \multirow{5}{1.4cm}{\centering Mantle} & Thermal expansivity ($\alpha$) & $3\cdot 10^{-5}$ K$^{-1}$ & B\\
           	 & Specific heat capacity ($C_p$) & 1250 J kg$^{-1}$ K$^{-1}$ & B C\\
             & Thermal conductivity ($\kappa$) & 6 W m$^{-1}$ K$^{-1}$ & B C\\
             & Grüneisen parameter ($\gamma$) & 1.45 & B C E\\
             & Homologous constant ($\xi$) & 12.3 & E\\
             \hline
     \end{tabular}
          \tablebib{
        (A)~\citet{Labrosse}; (B) \citet{Gaidos}; (C) \citet{Zuluaga13}; (D) \citet{Pozzo}; (E) \citet{Stamenkovic}; (F) This study.}
   \end{table*}

A key factor for defining the exoplanets thermal model is the core thermal conductivity. There are different values for this conductivity in the literature. For example, \citet{Gaidos} propose a core thermal conductivity of 35 W/m/K, \citet{Zuluaga13} a value of 40 W/m/K, \citet{Labrosse} a value in the range 50-60 W/m/K, and \citet{Pozzo} a value of 100 W/m/K. We have used our model, with the details explained in the Appendix, to the Earth, Venus, Mercury, Mars, and Ganymede and we have found that the value for the core thermal conductivity better fitting the observed magnetic moment and local Rossby number is 60 W/m/K, confirming the estimations of \citet{Labrosse}. With this value, the characteristics of the core and the magnetic moment can be calculated for these five objects. These values are summarized in Table \ref{magnetic}. In this table, we can see how our model describes properly the observed values.

\subsection{Magnetic field}

With the magnetic moment, it is possible to estimate the intensity of the dipolar magnetic field $B_{0dip}$ at $r_0$. It follows the relation

\begin{equation}
    B_{0dip}=\frac{\sqrt{2}\mu_0}{4\pi r_0^3}\mathcal{M}
\end{equation}

This value can be extrapolated to the planet surface ($B_{sdip}$) using \citep{Gaidos,Driscoll}

\begin{equation}
    B_{sdip}=B_{0dip}\Big(\frac{r_0}{R_p}\Big)^3
\end{equation}

\section{Application to the first set of exoplanets discovered by TESS }

To test how our model works on real data, we have applied our method to calculate the composition and magnetic moment of dry and water-rich rocky planets found at the first 176 planets confirmed by the TESS space mission. For doing that we have used the exoplanet's data available at TESS database on January 2022\footnote{\url{https://tess.mit.edu/publications/#list_of_tess_planets}}. In general, in this study the following types of exoplanets will be considered:

\begin{itemize}
    \item Dry Rocky exoplanets. Those exoplanets that, for a given mass, have a radius equal to or less than the radius producing a CMF=0 (See Table \ref{Grid}). They are mainly iron and silicate worlds where the presence of water is residual, as is the case of the Earth.
    \item Water-Rich Rocky exoplanets.  They are those exoplanets that, for a given mass, have a radius greater than the radius producing a CMF=0 and equal to or lower to the radius corresponding to an IMF=1. That is, from the radius corresponding to CMF=0, the decrease in density cannot be justified with an increase in the presence of silicates and a lighter element such as water is necessary. If an exoplanet of this type is in the HZ, we call it an “Ocean planet”.
    \item Ice Giants. They are worlds of ice and silicates in the core, and a gas envelope in the outer zone made mainly of Hydrogen and Helium, where the planetary core predominates. They are those planets that, for a given mass, have a radius greater than that making IMF=1 and lower to or equal to the radius corresponding to a mixture of 50$\%$ of H-He. That is, from the radius corresponding to IMF=1, the decrease in density cannot be justified with an increase in the presence of water and lighter elements such as Hydrogen and Helium are necessary. 
    \item Gas Giants. They are worlds of Ice and Silicates in the nucleus, and a gas envelope in the outer area of Hydrogen and Helium, where the core's envelope predominates. They are those planets that for a given mass has a radius greater than the corresponding to 50$\%$ of H-He. 
\end{itemize}

From a statistical point of view, dry rocky planets do not usually have a radius greater than 2 R$_\oplus$, while water-rich rocky planets are rare beyond 2.8 R$_\oplus$. Ice Giants range from 2.3 to 7 R$_\oplus$, while Gas Giants typically have more than 7 R$_\oplus$. 

The giant planets have not been analysed here as they are not the target of this study. For the rest of the planets, we use the next tools:

\begin{itemize}
    \item We assume a Gaussian distribution for the observational uncertainties. Therefore, we generate a set of 30000 sampling values of the planet mass, radius and rotational period following the observed values.
    \item For these 30000 sampling points we evaluate the probability of this planet to be one of the different types.
    \item When we don´t know exoplanet mass, it is estimated following \citep{Chen17}. In the TESS database, some planets are shown with a mass without uncertainties. If using this mass as input in our mass-radius diagram the planet appears above the composition of 100 $\%$ of Fe, we assume that this mass has no physical sense and we follow the procedure of the absence of known planetary mass.
    \item To distinguish whether the exoplanet is tidally locked, we use the equations of \cite{Griessmeier09}.
    \item For tidally locked exoplanets, from the orbital eccentricity we estimated the spin-orbit most likely relation following \citep{Dobrovolskis07}.
    \item To determine the mean effective flux of energy that a planet receives from its star, the eccentricity of the orbit has been taken into account \citep{Kopparapu13}.
    \item The HZ of each star has been determined following \citep{Kopparapu13}. The displacement of the inner edge of the HZ for synchronized rotating planets around low-mass stars has also been taken into account according to \citep{Kopparapu16}, as well as the effect of the planet mass on the HZ \citep{Kopparapu14}. The boundaries of the HZ ($D_2$ for the inner boundary and $D_3$ for the outer boundary) have been obtained as the mean values of the 30000 samples.
    \item  We will assume that all planetary dynamos are active since the TESS database does not include the Age of stars.
\end{itemize}

The results we obtain are summarized in Table \ref{TESS_dry} for dry rocky planets and Table \ref{TESS_water} for water-rich planets. In these tables we can see the situation of the planets with respect to the HZ ($D_2$ and $D_3$), the probability of being this kind of planet (Prob.), its composition (core mass fraction, mantle mass fraction, and ice mass fraction), and the magnetic properties (local Rossby number, regime and normalized magnetic moment).

For dry rocky planets, it is remarkable that all confirmed planets are tidally locked since the orbital periods are very small, and their spin-orbit resonance is 1:1, except LHS 1678 b and TOI-2285 b which have a spin-orbit resonance of 3:2. All these planets, except TOI-700 d, are in the hot zone of their star receiving a flow of energy making it impossible for water to be in the liquid phase. Approximately one-half of these planets have an Earth-like composition (CMF= $0.325  \pm 20\%$). Regarding the magnetic moment, half of these planets may have a powerful dipole magnetic moment, but they are in orbits so close to their star and they hardly can effectively protect the planet from possible violent events from their star. This result on which Mercury-like planets dominate may be an observational bias because the first discovered TESS exoplanets, those we analyse in this work, present short orbital periods.

The most interesting dry rocky planet so far, according to our analysis, is TOI-700d, which is the only one located of this class in the list at the HZ and it may contain liquid water on its surface. However, it has a slow rotation (37.42 days) and its magnetic moment is weak (0.01 $\mathcal{M}_\oplus$) but ten times larger than that of Mercury. Therefore, it is very exposed to the action of erosion from stellar winds. If it has been able to maintain its atmosphere, taking into account that it rotates in synchronous rotation with its star and the flow of energy is not much less than that received by the Earth from the Sun (0.87 $S_{\rm eff,\oplus}$), at the equator of the day face of the planet there could be an ocean of liquid water, although the polar ice caps could be larger than the terrestrial ones. 

When looking at the water-rich rocky planets (Table \ref{TESS_water}), it is only noteworthy that they are all tidally locked and in the hot zone of their star, except TOI-2257 b. According to our model it is an ocean planet located in the HZ of its star with a small core that could generate a magnetic field four times larger than that of Ganymede. Due to the absence of an observed mass for this planet, we have estimate it from models, following the procedure described above.

\section{Conclusions}

In this work, we have determined an internal structure model of rocky planets. We have defined three reference models: Earth's internal structure defined by PREM to be used for medium-high density dry rocky exoplanets, Mars assuming an Earth-like mantle composition for low-density dry rocky exoplanets, and Ganymede for water-rich rocky exoplanets. The main idea of our work is to extrapolate, in each case, the internal structure of one of the reference models for estimating the internal characteristics of both dry and water-rich rocky exoplanets and the similar objects in the solar system, these last used for testing the performance of our model.

The internal structure model we propose has some differences with previous proposals in the literature which can be summarized as:

\begin{itemize}
    \item The solid part of the core is not discarded. In the case of Earth, it means an error in mass larger than a 2$\%$.
    \item We take into account the possible existence of inner liquid oceans in water-rich exoplanets. We have also included the option of the presence of saltwater on them.
    \item To calculate the internal structure of water-rich planets it is imperative to use the moment of inertia factor. It allows the accurate determination of the beginning and depth of these liquid oceans, and, together with the imposed null error for the total mass and the corresponding EOS, allows the determination of the thicknesses of the ice, mantle, and core layers.
    \item We have also modeled the key thermodynamic variables as a function of the internal pressure.
\end{itemize}

With this internal structure model, a new Mass-Radio diagram has been constructed based on extrapolating the three reference models.

The Magnetic Moment is a fundamental factor protecting the atmosphere of an exoplanet, and therefore its potential life on its surface, from the erosion of stellar winds and cosmic rays \citep{Mozos19}. To determine the magnetic moment of an exoplanet and if it is dipolar or multipolar we have used the scaling laws of planetary dynamos of \citep{Olson06}. For extrapolating these scaling laws to exoplanets more massive than Earth, both dry and water-rich, different works have been used to obtain the variation with the pressure of the thermodynamic parameters involved in these planetary dynamos.

Finally, the proposed procedure has been applied to the first 176 exoplanets confirmed by TESS. Up to our knowledge, it is the first time that a massive estimation of the magnetic properties of exoplanets is done. We have presented the situation of these exoplanets in the HZ of their hosting stars. For all those rocky planets in the list, we have obtained their composition and magnetic properties assuming that this dynamo is still active. We have found that the most interesting objects, from an astrobiological point of view, are TOI-700 d and TOI-2257 b.

%

%

\begin{acknowledgements}
The authors want to thank the referee, Peter Olson, and the editor for their very constructive comments that have certainly improved this manuscript. AM acknowledges funding support from Grant PID2019-107061GB-C65 funded by MCIN/AEI/ 10.13039/501100011033, and from Generalitat Valenciana in the frame of the GenT Project CIDEGENT/2020/036.
\end{acknowledgements}

%
%

\newpage
\begin{appendix}
\section{Thermal model}
A key element in determining a planet's magnetic moment is the buoyancy convective flux that is related to the  heat flow generated by the core. To estimate both variables it is necessary to have a thermal model. The model we have constructed is described below.

In this study, we will follow the thermal models made for the Earth by of \citep{Labrosse}, \citep{Gaidos}, and \citep{Zuluaga13} for, afterwords, extrapolate the results to rocky exoplanets that, in general, are more massive and have higher interior temperatures and pressures. For example, the pressure at the CMB of the Earth reaches up to 135 GPa, but in rocky exoplanets, this pressure can reach up to TPa. This has a large impact on the interior heat transport models, and thermal properties such as thermal expansivity, specific heat, or viscosity must be modeled at extreme pressure conditions \citep{Stamenkovic}.

For doing so, we firstly are going to estimate the temperature gradient of the exoplanets. With this gradient, we can describe the heat extracted from the core. 

\subsection{Temperature gradient}

Assuming a steady-state for the exoplanet, if we know the adiabatic gradient of temperatures we can estimate the temperature at any position at the mantle. This can be done using \citep{Labrosse}

\begin{equation}
    T(r)=T_m\cdot e^{\Big(\frac{R_p^2-r^2}{D_m^2}\Big)}
\end{equation}

\noindent where $T_m$ is the temperature in the outermost part of the Upper Mantle, and $D_m$ is the temperature high scale of the mantle that can be calculated as \citep{Labrosse}

\begin{equation}
    D_m=\sqrt{\frac{3C_p}{2\pi \gamma \alpha \rho {\rm G}}}
    \label{dm_mantle}
\end{equation}

\noindent being $\gamma$ the Grüneisen parameter, and G the gravitational constant. In this equation, $C_p$, $\alpha$, and $\rho$ are those at the bottom of the mantle, at the boundary with the CMB. Using these two equations, the temperature at this bottom limit of the mantle ($T_\ell$) can be estimated.

For calculating the temperature at the beginning of the core ($T_c$) we can assume that the temperature gradient in this zone ($\Delta T_{\rm CMB}$) is proportional to the temperature increment produced at the mantle \citep{Zuluaga13}, that is

\begin{equation}
    \Delta T_{\rm CBM}=\epsilon_{\rm ad}(T_\ell-T_m)  
\end{equation}

\noindent where $\epsilon_{\rm ad}$ is the adiabatic factor of the CMB. \citet{Zuluaga13} estimated that the value for this adiabatic factor better reproducing the Earth is $\epsilon_{\rm ad}=0.7$. Therefore

\begin{equation}
    T_c=T_\ell+\Delta T_{\rm CMB} 
\end{equation}

In the core we can also assume an adiabatic gradient of temperatures

\begin{equation}
    T(r)=T_c\cdot e^{\Big(\frac{r_0^2-r^2}{D_c^2}\Big)}
    \label{grad_core}
\end{equation}

\noindent where $D_c$ is the temperature high scale of the core that can be calculated as \citep{Labrosse} 

\begin{equation}
    D_c=\sqrt{\frac{3C_p}{2\pi \alpha \rho {\rm G}}}
\end{equation}

The only difference of this equation with equation \ref{dm_mantle} is the absence of the Grüneisen parameter. Finally the temperature at the planet centre ($T_{cc}$) is estimated using Equation \ref{grad_core} at $r=0$.

To fix the temperature profile of the planet it is necessary to know the average temperature of the mantle $T_{mm}$, which can be calculated using the following equation \citep{Gaidos} 

\begin{equation}
    T_{mm} =\theta \cdot e^{\frac{1}{2}\int_{r_0}^{R_p} \frac{\alpha g} {C_p} dr}
\end{equation}

\noindent being $\theta$ the potential temperature of the mantle we will take equal to 1700 K \citep{Zuluaga13}

Therefore, for estimating the temperature profile of an exoplanet, in general more massive than Earth, we need to know how $C_p$, $\alpha$, $\gamma$, and $\rho$ change with pressure. The density profile is provided by our internal structure model, and the other three variables are taken from \citep{Stamenkovic}.

In water-rich exoplanets, the temperature gradient can be known from the water melt curve of the different crystallization systems via the EOS described in Section \ref{model_wr}. With this information, we can estimate the pressure and temperature at the boundary ice layer - mantle and use the equations here described to obtain the temperature for the rest of the exoplanet.

\subsection{Heat extracted from the core}

The CMB is a relatively narrow zone. \citep{Driscoll} estimated that it can have a thickness up to 286 km with a temperature gradient of 5.5 K/km. On the other hand, \citep{Okuda} estimated a thickness of 200 km and a temperature gradient of 7 K/km. The planetary dynamo activity is related to the heat transport and the temperature gradient at CMB \citep{Gaidos}. Heat transport in the Earth's interior occurs predominantly by convection. However, heat transport in the CMB occurs only by conduction \citep{Okuda}.

The total heat released by the core ($Q_c$) can be obtained using the approximation shown in \citep{Ricard}

\begin{equation}
    Q_c=4\pi r_0 \kappa_m(T_c-T_\ell)Nu_c
\end{equation}

\noindent where $\kappa_m$ is the thermal conductivity of the lower mantle and $Nu_c$ is the Nusselt number at the core and it is defined as

\begin{equation}
    Nu_c=\Big(\frac{Ra_c}{Ra_*}\Big)^\delta
\end{equation}

\noindent being $Ra_*$ the critical Rayleigh number, and $Ra_c$ this number at the CMB. For our model, we have adopted the values proposed by \citep{Gaidos}, that is, $Ra_*=1100$ and $\delta=0.3$. $Ra_c$ can be calculated assuming a CMB that is heated from below and using the equation \citep{Ricard}

\begin{equation}
    Ra_c=\frac{\rho g \alpha (T_c-T_\ell)(R_p-r_0)^3}{\lambda_t\eta_c}
\end{equation}

\noindent where $\eta_c$ is the dynamic viscosity at the CMB.

\citet{Manga} showed that the temperature gradients at the CMB are produced at two narrow zones of this layers close to the core and mantle boundaries respectively. At the rest of the CMB the temperature remains almost constant. This constant value ($T_{\rm CMB}$) can be estimated as a value between the upper core temperature and the bottom mantle temperature, that is

\begin{equation}
    T_{\rm CMB}=\theta_cT_c+(1-\theta_c)T_\ell
\end{equation}

\noindent where $\theta_c$ is inversely proportional to the ratio between the lower mantle and core viscosities, that is, its value must be lower than 0.5 \citep{Manga}. For our model, we have used the value proposed by \citep{Zuluaga13} for Earth of $\theta_c=0.4$. 

Another variable we must estimate is the dynamical viscosity at the CMB ($\eta_c$). The dynamical viscosity of a mineral at large pressures and temperatures can be obtained using the Navarro-Herring model \citep{Yamazaki} with the following expression

\begin{equation}
    \eta_c=\frac{R_gd^m}{D_0Am_{\rm mol}}\rho T_{\rm  CMB}\cdot e^{\Big(\xi\frac{T_{\rm melt}}{T_{\rm CMB}}\Big)}
\end{equation}

\noindent with $R_g$ the ideal gas constant in mols, $d$ the grain size, $m$ the grain growth rate, $D_0$ the pre-exponential diffusion coefficient, $A$ the viscosity pre-exponential coefficient, $m_{\rm mol}$ the molar weight, $\xi$ the homologous constant, and $T_{\rm melt}$ the melting temperature of the constituent. For our model, we have used the values displayed in Table 1 of \citep{Stamenkovic} (see Table \ref{eath_term}). For $T_{\rm melt}$ a good approximation is using the following fifth-order polynomial

\begin{equation}
    T_{\rm melt}=\sum_{i=1}^5{a_iP^i}
\end{equation}

With the following coefficients: ($a_0$, $a_1$, $a_2$, $a_3$, $a_4$, $a_5$)=(2752.4, 22.817, -0.013104, 8.8756$\cdot 10^{-6}$, -3.02732$\cdot 10^{-9}$, 3.9362$\cdot 10^{-13}$), \citep{Stamenkovic} found an inaccuracy of this approximation lower than a $1.5\%$ above 25 GPa.

The last ingredient to determine the heat released by the core is the thermal conductivity of the lower mantle ($\kappa_m$) which is one of the most important properties to understand the thermal dynamics in the CMB. Recent observations of thermal conductivity of Fe$_{\rm x}$ Mg$_{\rm 1-x}$ O show an important reduction of its value due to the presence of Fe impurities \citep{Ohta}. As we have already said, pv transform into ppv at a pressure of 125 GPa and a temperature of 2500 K. These reduction impacts mainly on ppv and not on pv \citep{Okuda}. To determine the variation of thermal conductivity with the pressure at ambient temperature (300 K), the measures for pv have been used \citep{Ohta}, while for pressures larger than 125 Gpa we have used a mix of ppv with a 3$\%$ of Fe \citep{Okuda} and impurities of Fe$_{\rm x}$ Mg$_{\rm 1-x}$ O.

This thermal conductivity at ambient temperature, which we will call $\kappa_{m0}$, can be extrapolated to the pressure and temperatures conditions at the CMB, using the density-temperature model described by \citep{Okuda}

\begin{equation}
    {\kappa_m}=\kappa_{m0}\Big(\frac{\rho}{\rho_0}\Big)^{e_1}\Big(\frac{T_0}{T}\Big)^{e_2}    
\end{equation}

In this expression, $T_0=300$ K, $\rho_0=5470$ kg/m$^3$, and the exponents are presented in \citep{Okuda} with values $e_1=6\pm 0.3$ and $e_2=0.65\pm 0.04$. The results for $\kappa_m$ at the pressure and temperature conditions of the CMB we obtain are very similar to those presented by \citep{Gaidos} and \citep{Zuluaga13} of $\kappa_m=6$ W/m/K. These results have been recently confirmed by thermal conductivity measurements of ppv+Fe$_{\rm x}$ Mg$_{\rm 1-x}$ O at 124 GPa and temperatures between 2000 and 3000 K, with a value of 5.9 W/m/K \citep{Geballe}.

In the case of water-rich rocky planets, for the same mass and similar core size, $Q_c$ is lower thanks to lower pressure and temperature conditions at the CMB compared with the case of a dry rocky planet.

\section{Long tables}

\longtab[1]{
\begin{landscape}
\begin{longtable}{c|cccccccc|cccccccc}
 \caption[]{\label{Grid} Grid Mass-Radius for Rocky exoplanets.}\\
    
$M_p$ & R100  &  R50 &  R40   &    R30   &    R25   &    R20   &    R10    &   R0  & H10   &    H20    &   H30   &    H40    &   H50   &    H60    &   H80   &    H100\\
    \hline
  0.01   &  0.197  &   0.225   &  0.237  &   0.241  &   0.243   &  0.246  &   0.250  &   0.254  &   0.269&0.283&0.295&0.306&0.317&0.327&0.345&0.362\\
  0.05&0.332&0.389&0.402&0.410&0.414&0.418&0.425&0.432&0.450&0.471&0.490&0.506&0.522&0.535&0.563&0.588\\
  0.10&0.414&0.487&0.504&0.514&0.519&0.523&0.533&0.542&0.562&0.587&0.608&0.627&0.645&0.662&0.694&0.724\\
  0.25&0.550&0.647&0.661&0.681&0.686&0.692&0.703&0.714&0.741&0.771&0.798&0.823&0.847&0.869&0.912&0.950\\
  0.50&0.675&0.790&0.808&0.825&0.833&0.842&0.860&0.873&0.904&0.941&0.973&1.003&1.030&1.057&1.111&1.162\\
  1.0& 0.822&0.961&0.984&1.005&1.016&1.026&1.047&1.067&1.102&1.145&1.183&1.221&1.256&1.289&1.351&1.412\\
  1.5& 0.917&1.076&1.101&1.126&1.138&1.150&1.174&1.197&1.235&1.281&1.327&1.368&1.407&1.443&1.511&1.577\\
  2.0& 0.990&1.163&1.191&1.219&1.232&1.246&1.272&1.298&1.337&1.390&1.438&1.481&1.522&1.561&1.634&1.705\\
  2.5& 1.048&1.244&1.272&1.298&1.311&1.323&1.347&1.378&1.425&1.487&1.539&1.585&1.627&1.665&1.735&1.799\\
  3.0& 1.096&1.295&1.328&1.359&1.375&1.390&1.420&1.450&1.492&1.553&1.605&1.652&1.697&1.739&1.821&1.900\\
  4.0& 1.176&1.395&1.431&1.466&1.483&1.500&1.533&1.567&1.612&1.676&1.731&1.782&1.830&1.876&1.965&2.049\\
  6.0& 1.292&1.544&1.585&1.625&1.644&1.664&1.703&1.741&1.791&1.860&1.921&1.978&2.031&2.083&2.183&2.279\\
  8.0& 1.378&1.665&1.700&1.744&1.766&1.787&1.830&1.872&1.925&1.999&2.064&2.126&2.185&2.242&2.351&2.456\\
  10.0&1.446&1.744&1.792&1.840&1.863&1.886&1.932&1.978&2.033&2.111&2.181&2.246&2.310&2.370&2.486&2.598\\
  12.0&1.503&1.818&1.870&1.920&1.944&1.969&2.018&2.066&2.124&2.205&2.279&2.349&2.415&2.478&2.598&2.715\\
  16.0&1.594&1.939&1.995&2.050&2.076&2.104&2.157&2.211&2.271&2.359&2.440&2.515&2.584&2.649&2.776&2.902\\
  20.0&1.665&2.035&2.095&2.154&2.183&2.211&2.269&2.326&2.389&2.484&2.568&2.645&2.716&2.784&2.915&3.064\\
  24.0&1.722&2.115&2.178&2.240&2.270&2.300&2.360&2.421&2.487&2.587&2.674&2.751&2.824&2.894&3.029&3.164\\
  32.0&1.810&2.240&2.308&2.374&2.407&2.439&2.505&2.571&2.643&2.749&2.838&2.918&2.994&3.067&3.209&3.350\\
  40.0&-&-&-&-&-&-&-&-&2.756&2.867&2.960&3.045&3.123&3.199&3.347&3.495\\
  48.0&-&-&-&-&-&-&-&-&2.841&2.952&3.048&3.137&3.221&3.303&3.459&3.614\\
  \hline

\end{longtable}
\tablefoot{$M_p$ represents the planetary mass in M$_\oplus$, R* is the exoplanet radius (in R$_\oplus$) for a CMF=1.00, 0.5, 0.4, 0.3, 0.25, 0.2, 0.1, and 0, respectively. H* is the exoplanet radius (in R$_\oplus$) for an IMF= 0.1, 0.2, 0.3, 0.4, 0.5, 0.8, and 1.0, respectively. For masses larger than 1M$_\oplus$ we extrapolate the Earth's internal structure using the proper thermal conductivity and energy flow adapted to that mass.}
   \end{landscape}
   }

\longtab[2]{
\begin{landscape}
\begin{longtable}{c|ccc|ccc|cc|ccc} 
\caption{\label{magnetic} Core Characteristics and Magnetic Properties of planets and satellites of Solar System.}\\
\hline
&\multirow{2}{2.4cm}{Normalized Planet Mass}	& \multirow{2}{2.4cm}{Normalized Planet Radius}	& Rotational Period	& \multicolumn{3}{c|}{Core Characteristics}	& \multicolumn{5}{c|}{Magnetic Properties}\\
\\
\hline
Planet	& $M_p$	& $R_p$ &	days &	$\frac{\rho_0}{\rho_{0\oplus}}$	& $\frac{r_0}{r_{0\oplus}}$ &	$\frac{F}{F_\oplus}$ &	\multirow{2}{1.2cm}{Calculated $R_{o\ell}$} &	\multirow{2}{1.2cm}{Observed $R_{o\ell}$}	& \multirow{2}{1.2cm}{Calculated $\mathcal{M}/\mathcal{M}_\oplus$}	& \multirow{2}{1.2cm}{Observed $\mathcal{M}/\mathcal{M}_\oplus$} &	\multirow{2}{1.2cm}{Error $\mathcal{M}/\mathcal{M}_\oplus$}\\
\\
\hline
Mercury	& 0.0533 (1) &	0.3829 (2) &	58.8 (3) &	0.70	& 0.55	& 0.32	& 8.1	& 8 (3)	& 0.0003	& 0.0004 (3)	& 0.001\\
Venus	& 0.815 (4)	& 0.9499 (2)	&243.7 (3)	&0.96	&0.92	&0.89	&50.7	&50 (3)	&0.0007	&0 (3)	& 0.001\\
Earth	&1 (5)	&1 (2)	&1 (3)	&1.00	&1.00	&1.00	&0.09	&0.09 (3)	&1.00	&1 (3)	& 0.000\\
Mars (a)	&0,1074 (6)	&0.532 (2)	&1 (3)	&0.64	&0.50	&0.94	&0.10	&0.10 (3)	&0.084	&0.10 (3)	&0.016\\
Ganymede	&0.0248 (7)	&0.4132 (2)	&0.73 (3)	&0.59	&0.22	&0.29	&0.05	&0.05 (3)	&0.003	&0.002 (3)	&0.001\\
\hline
\end{longtable}
    \begin{tablenotes}
        \item References: (a)~ Extinct dynamo;
        (1)~\citet{Smith12}; (2)~\citet{Seidelmann07}; (3)~\citet{Olson06}; (4)~\citet{Zeng16}; (5)~\citet{Dzie81}; (6)~\citet{Konopliv11}; (7)~\citet{Schubert04}.
        \end{tablenotes}
\end{landscape}
}
   
\longtab[3]{
\begin{landscape}
\setlength\tabcolsep{5.3pt}
     \begin{longtable}{c|cccc|cccc|ccc|ccc}
    \caption[]{\label{TESS_dry} Result for dry rocky planets observed by TESS after 30,000 runs (Monte Carlo method). }\\
\hline
      Exoplanet & $M_p$ & $R_p$ & S:O & $P_{\rm rot}$ & a & $D_2$ & $D_3$ & $S_{\rm eff}$ & Prob. & CMF & MMF & $R_{o\ell}$ & Regime & 
 $\mathcal{M}/\mathcal{M}_\oplus$\\
\hline
\endfirsthead
\caption{continued.}\\
\hline
      Exoplanet & $M_p$ & $R_p$ & S:O & $P_{\rm rot}$ & a & $D_2$ & $D_3$ & $S_{\rm eff}$ & Prob. & CMF & MMF & $R_{o\ell}$ & Regime & 
 $\mathcal{M}/\mathcal{M}_\oplus$\\
\hline
\endhead
\hline
\endfoot
GJ 1252 b   & 2.119±0.311 & 1.190±0.042 & 1:1 & 0.518 & 0.009 & 0.144 & 0.282 & 234.7 & 0.94 & 0.47±0.20 & 0.53±020 & 0.03±0.01 & Dipolar & 3.50±1.40\\
GJ 3473 b   & 1.965±0.135 & 1.252±0.025 & 1:1 & 1.198 & 0.016 & 0.124 & 0.245 & 59.2 & 0.54 & 0.15±0.10 & 0.85±0.10 & 0.14±0.03 & Multip. & 0.05±0.02\\
GJ 357 b    & 1.884±0.171 & 1.202±0.043 & 1:1 & 3.931 & 0.034 & 0.129 & 0.252 & 13.5 & 0.77 & 0.29±0.16 & 0.71±0.16 & 0.47±0.09 & Multip. & 0.04±0.02\\
GJ 367 b    & 0.546±0.045 & 0.718±0.031 & 1:1 & 0.322 & 0.007 & 0.182 & 0.343 & 588.2 & 1.00 & 0.92±0.015 & 0.08±0.015 & 0.01±0.002 & Dipolar & 0.86±0.38\\
HD 15337 b  & 7.547±0.607 & 1.640±0.035 & 1:1 & 4.756 & 0.053 & 0.644 & 1.201 & 160.9 & 1.00 & 0.48±0.10 & 0.52±0.10 & 0.37±0.07 & Multip. & 0.33±0.15\\
HD 213885 b & 8.835±0.379 & 1.744±0.030 & 1:1 & 1.008 & 0.020 & 1.067 & 1.939 & 3420 & 1.00 & 0.39±0.07 & 0.61±0.07 & 0.07±0.01 & Dipolar & 12.59±1.50\\
HD 21749 c  & 0.670±0.038$^*$ & 0.886±0.031 & 1:1 & 7.790 & 0.069 & 0.458 & 0.823 & 41.4 & 0.84 & 0.36±0.16 & 0.64±0.16 & 1.03±0.21 & Multip. & 0.01±0.002\\
L 168-9 b   & 4.600±0.323 & 1.390±0.052 & 1:1 & 1.402 & 0.021 & 0.259 & 0.512 & 157.7 & 1.00 & 0.64±0.13 & 0.36±0.13 & 0.04±0.01 & Dipolar & 9.11±1.09\\
LHS 1478 b  & 2.330±0.116 & 1.243±0.029 & 1:1 & 1.950 & 0.019 & 0.088 & 0.173 & 20.3 & 1.00 & 0.40±0.11 & 0.60±0.11 & 0.15±0.03 & Multip. & 0.17±0.08\\
LHS 1678 b  & 0.350±0.000 & 0.696±0.025 & 3:2 & 0.573 & 0.012 & 0.130 & 0.242 & 98.0 & 1.00 & 0.60±0.13 & 0.40±0.13 & 0.05±0.01 & Dipolar & 0.64±0.03\\
LHS 1678 c  & 1.400±0.000 & 0.983±0.037 & 1:1 & 3.694 & 0.032 & 0.124 & 0.242 & 13.9 & 1.00 & 0.77±0.12 & 0.23±0.12 & 0.15±0.03 & Multip. & 0.14±0.01\\
LHS 1815 b  & 1.356±0.077$^*$ & 1.085±0.036 & 1:1 & 3.814 & 0.038 & 0.205 & 0.396 & 27.6 & 0.96 & 0.34±0.16 & 0.66±0.16 & 0.42±0.08 & Multip. & 0.04±0.02\\
LHS 3844 b  & 2.295±0.127$^*$ & 1.302±0.012 & 1:1 & 0.463 & 0.006 & 0.054 & 0.108 & 71.2 & 0.95 & 0.17±0.08 & 0.83±0.08 & 0.04±0.01 & Dipolar & 1.53±0.69\\
LP 791-18 b & 1.508±0.086$^*$ & 1.093±0.061 & 1:1 & 0.948 & 0.010 & 0.046 & 0.093 & 20.8 & 0.79 & 0.44±0.23 & 0.54±0.23 & 0.07±0.01 & Dipolar & 2.01±0.96\\
LTT 3780 b  & 2.681±0.257 & 1.323±0.040 & 1:1 & 0.768 & 0.012 & 0.126 & 0.250 & 103.8 & 0.86 & 0.29±0.15 & 0.71±0.15 & 0.06±0.01 & Dipolar & 2.97±1.34\\
TOI-1235 b  & 5.992±0.342 & 1.677±0.038 & 1:1 & 3.445 & 0.038 & 0.294 & 0.577 & 60.1 & 0.72 & 0.17±0.10 & 0.83±0.10 & 0.42±0.08 & Multip. & 0.10±0.04\\
TOI-1444 b  & 3.871±0.408 & 1.397±0.037 & 1:1 & 0.470 & 0.012 & 0.760 & 1.388 & 4838 & 1.00 & 0.46±0.14 & 0.54±0.14 & 0.03±0.01 & Dipolar & 6.59±1.32\\
TOI-1749 c  & 14.00±0.00 & 2.069± 0.040 & 1:1 & 4.493 & 0.044 & 0.260 & 0.509 & 34.7 & 0.58 & 0.14±0.08 & 0.86±0.08 & 0.57±0.11 & Multip. & 0.15±0.01\\
TOI-178 b   & 1.589±0.196 & 1.141±0.039 & 1:1 & 1.915 & 0.026 & 0.368 & 0.689 & 192.6 & 0.67 & 0.31±0.17 & 0.69±0.17 & 0.20±0.04 & Multip. & 0.07±0.03\\
TOI-2285 b  & 19.50±0.00 & 1.740±0.046 & 3:2 & 18.180 & 0.137 & 0.168 & 0.338 & 1.6 & 1.00 & 0.92±0.06 & 0.08±0.06 & 0.48±0.10 & Multip. & 0.57±0.03\\
TOI-270 b   & 1.677±0.109 & 1.198±0.021 & 1:1 & 3.360 & 0.032 & 0.144 & 0.281 & 19.0 & 0.56 & 0.15±0.09 & 0.85±0.09 & 0.46±0.09 & Multip. & 0.02±0.01\\
TOI-431 b   & 3.069±0.202 & 1.280±0.023 & 1:1 & 0.490 & 0.011 & 0.504 & 0.935 & 2112 & 1.00 & 0.57±0.08 & 0.43±0.08 & 0.02±0.004 & Dipolar & 5.86±1.17\\
TOI-540 b   & 0.699±0.039$^*$ & 0.899±0.028 & 1:1 & 1.239 & 0.012 & 0.062 & 0.120 & 23.1 & 0.90 & 0.35±0.15 & 0.65±0.15 & 0.12±0.02 & Dipolar & 0.75±0.30\\
TOI-700 b   & 1.143±0.065$^*$ & 1.034±0.036 & 1:1 & 9.977 & 0.067 & 0.156 & 0.304 & 5.0 & 0.95 & 0.34±0.16 & 0.66±0.16 & 1.30±0.26 & Multip. & 0.02±0.01\\
TOI-700 d   & 1.620±0.092$^*$ & 1.144±0.035 & 1:1 & 37.425 & 0.163 & 0.155 & 0.304 & 0.87 & 0.98 & 0.33±0.15 & 0.67±0.15 & 5.66±0.94 & Multip. & 0.01±0.002\\
\hline
     \end{longtable}
     \tablefoot{$M_p$ is the planetary mass normalized to the Earth mass. Those values highlighted with an asterisk have been estimated following \citep{Chen17}. $R_p$ is the planetary radius normalized to the Earth radius.  S:O shows the Spin:Orbit resonance. $P_{\rm rot}$  is the planetary rotational period in days. $a$, $D_2$ and $D_3$ are respectively are the semi-major axis, inner and outer boundaries of the HZ expressed in A.U. $S_{\rm eff}$ is normalized mean effective flux considering orbital eccentricity. {\it Prob.} It is the probability of being rocky planet. All the planets are tidally locked. Errors in the rotational period have not been included because they are so small that they have no impact on the magnetic properties of exoplanets.}
   \end{landscape}
}

\footnotesize
\longtab[4]{
\begin{landscape}
\setlength\tabcolsep{4.3pt}
\begin{longtable}{c|cccc|cccc|cccc|ccc}
\caption[]{\label{TESS_water} Result for water-rich rocky planets observed by TESS after 30,000 runs (Monte Carlo method).}\\
\hline
Exoplanet & $M_p$ & $R_p$ & S:O & $P_{\rm rot}$ & a & $D_2$ & $D_3$ & $S_{\rm eff}$ & Prob. & CMF & MMF & IMF & $R_{o\ell}$ & Regime & $\mathcal{M}/\mathcal{M}_\oplus$\\
\hline
\endfirsthead
\caption{continued.}\\
\hline
Exoplanet & $M_p$ & $R_p$ & S:O & $P_{\rm rot}$ & a & $D_2$ & $D_3$ & $S_{\rm eff}$ & Prob. & CMF & MMF & IMF & $R_{o\ell}$ & Regime & $\mathcal{M}/\mathcal{M}_\oplus$\\
\hline
\endhead
\hline
\endfoot
GJ 143 b    & 22.884±1.186 & 2.615±0.095 & 1:1 & 35.613 & 0.191 & 0.433 & 0.823 & 5.6 & 1.00 & 0.11±0.02 & 0.63±0.10 & 0.26±0.12 & 5.18±0.47 & Multip. & 0.04±0.01\\
HD 108236 b & 3.290±0.190$^*$ & 1.615±0.029 & 1:1 & 3.796 & 0.045 & 0.792 & 1.429 & 345.3 & 0.99 & 0.11±0.01 & 0.65±0.06 & 0.24±0.07 & 0.48±0.05 & Multip. & 0.04±0.01\\
HD 108236 c & 5.020±0.288$^*$ & 2.071±0.030 & 1:1 & 6.203 & 0.063 & 0.780 & 1.429 & 179.3 & 0.99 & 0.03±0.01 & 0.20±0.08 & 0.77±0.09 & 0.70±0.07 & Multip. & 0.01±0.002\\
HD 110113 b & 4.630±0.346 & 2.026±0.060 & 1:1 & 2.541 & 0.036 & 0.881 & 1.606 & 690.0 & 0.76 & 0.03±0.02 & 0.20±0.11 & 0.77±0.13 & 0.25±0.03 & Multip. & 0.02±0.003\\
HD 15337 c  & 8.522±0.903 & 2.371±0.065 & 1:1 & 17.178 & 0.126 & 0.644 & 1.201 & 29.0 & 0.75 & 0.03±0.02 & 0.20±0.11 & 0.77±0.13 & 2.39±0.24 & Multip. & 0.01±0.002\\
HD 207897 b & 14.396±0.925 & 2.503±0.046 & 1:1 & 16.202 & 0.117 & 0.573 & 1.071 & 26.6 & 1.00 & 0.08±0.01 & 0.45±0.08 & 0.47±0.09 & 2.56±0.26 & Multip. & 0.03±0.01\\
HD 23472 c  & 10.315±3.781 & 2.203±0.182 & 1:1 & 29.652 & 0.171 & 0.469 & 0.891 & 8.0 & 0.59 & 0.10±0.04 & 0.59±0.22 & 0.31±0.25 & 5.42±0.54 & Multip. & 0.02±0.003\\
HD 63433 b  & 5.413±0.300$^*$ & 2.130±0.050 & 1:1 & 7.108 & 0.072 & 0.802 & 1.470 & 144.6 & 0.77 & 0.03±0.02 & 0.16±0.09 & 0.81±0.11 & 0.76±0.08 & Multip. & 0.01±0.002\\
HD 86226 c  & 7.285±0.666 & 2.160±0.046 & 1:1 & 3.984 & 0.049 & 0.986 & 1.798 & 480.4 & 1.00 & 0.07±0.02 & 0.39±0.11 & 0.54±0.13 & 0.50±0.05 & Multip. & 0.04±0.01\\
HIP 97166 b & 19.996±0.868 & 2.740±0.075 & 1:1 & 10.289 & 0.089 & 0.644 & 1.199 & 58.6 & 1.00 & 0.07±0.02 & 0.40±0.10 & 0.53±0.12 & 1.42±0.14 & Multip. & 0.05±0.01\\
HR 858 b    & 5.094±0.286$^*$ & 2.085±0.037 & 3:2 & 2.391 & 0.048 & 1.350 & 2.440 & 1037 & 0.95 & 0.03±0.01 & 0.18±0.08 & 0.79±0.10 & 0.23±0.02 & Multip. & 0.02±0.003\\
HR 858 c    & 4.484±0.259$^*$ & 1.939±0.040 & 1:1 & 5.973 & 0.067 & 1.354 & 2.440 & 510.3 & 1.00 & 0.06±0.02 & 0.33±0.09 & 0.61±0.10 & 0.75±0.08 & Multip. & 0.02±0.003\\
HR 858 d    & 5.482±0.298$^*$ & 2.149±0.043 & 3:2 & 7.487 & 0.103 & 1.348 & 2.440 & 224.9 & 0.76 & 0.02±0.01 & 0.14±0.08 & 0.84±0.09 & 0.76±0.08 & Multip. & 0.01±0.002\\
L 98-59 b   & 0.371±0.076 & 0.864±0.030 & 1:1 & 2.253  & 0.022 & 0.115 & 0.215 & 24.4 & 0.74 & 0.12±0.02 & 0.68±0.15 & 0.20±0.17 & 0.20±0.02 & Multip. & 0.01±0.002\\
L 98-59 c   & 2.204±0.144 & 1.407±0.044 & 1:1 & 3.691 & 0.030 & 0.109 & 0.215 & 12.6 & 0.85 & 0.12±0.01 & 0.72±0.07 & 0.16±0.08 & 0.44±0.04 & Multip. & 0.03±0.01\\
L 98-59 d   & 1.942±0.161 & 1.531±0.062 & 1:1 & 7.451 & 0.048 & 0.110 & 0.215 & 4.9 & 0.99 & 0.06±0.03 & 0.38±0.16 & 0.56±0.18 & 1.10±0.11 & Multip. & 0.01±0.002\\
LTT 1445A b & 1.835±0.588 & 1.399±0.071 & 3:2 & 3.573 & 0.038 & 0.097 & 0.190 & 6.4 & 0.42 & 0.10±0.04 & 0.61±0.23 & 0.29±0.26 & 0.52±0.05 & Multip. & 0.02±0.003\\
LTT 3780 c  & 8.787±0.822 & 2.302±0.088 & 1:1 & 12.252 & 0.077 & 0.123 & 0.250 & 2.6 & 0.97 & 0.06±0.03 & 0.33±0.15 & 0.61±0.18 & 1.83±0.18 & Multip. & 0.02±0.003\\
TOI-1062 b  & 10.139±0.478 & 2.268±0.054 & 1:1 & 4.113 & 0.049 & 0.672 & 1.245 & 214.9 & 1.00 & 0.08±0.01 & 0.49±0.08 & 0.43±0.09 & 0.50±0.05 & Multip. & 0.08±0.01\\
TOI-1260 b  & 8.618±0.813 & 2.346±0.056 & 1:1 & 3.127 & 0.036 & 0.343 & 0.665 & 91.1 & 0.95 & 0.04±0.02 & 0.25±0.12 & 0.71±0.13 & 0.35±0.05 & Multip. & 0.04±0.01\\
TOI-1266 b  & 14.874±4.645 & 2.397±0.079 & 1:1 & 10.895 & 0.074 & 0.163 & 0.326 & 4.9 & 0.74 & 0.10±0.04 & 0.60±0.22 & 0.31±0.26 & 1.61±0.16 & Multip. & 0.06±0.01\\
TOI-1266 c	& 2.695±0.756 & 1.575±0.078 & 1:1 & 18.801 & 0.106 & 0.167 & 0.326 & 2.4 & 0.65 & 0.10±0.04 & 0.57±022 & 0.33±0.26 & 3.03±0.30 & Multip. & 0.01±0.002\\
TOI-1634 b	& 4.899±0.397 & 1.790±0.046 & 1:1 & 0.989 & 0.015 & 0.170 & 0.341 & 122.9 & 1.00 & 0.11±0.01 & 0.65±0.09 & 0.24±0.10 & 0.10±0.01 & Dipolar & 2.54±0.51\\
TOI-1685 b	& 3.782±0.364 & 1.700±0.041 & 1:1 & 0.669 & 0.012 & 0.175 & 0.349 & 214.9 & 1.00 & 0.10±0.02 & 0.61±0.10 & 0.29±0.11 & 0.06±0.003 & Dipolar & 1.67±0.33\\
TOI-1749 b	& 2.536±0.146$^*$ & 1.492±0.064 & 1:1 & 2.388 & 0.029 & 0.265 & 0.509 & 81.3 & 0.53 & 0.11±0.02 & 0.67±0.10 & 0.22±0.12 & 0.28±0.03 & Multip. & 0.04±0.01\\
TOI-1749 d	& 15.00±0.00 & 2.520±0.086 & 1:1 & 9.051 & 0.071 & 0.259 & 0.509 & 13.7 & 1.00 & 0.08±0.02 & 0.46±0.11 & 0.46±0.13 & 1.29±0.13 & Multip. & 0.05±0.01\\
TOI-178 c	& 4.634±0.348 & 1.703±0.050 & 1:1 & 3.238 & 0.037 & 0.358 & 0.689 & 95.5 & 0.73 & 0.12±0.01 & 0.74±0.07 & 0.14±0.08 & 0.41±0.04 & Multip. & 0.07±0.01\\
TOI-1807 b	& 4.138±0.239$^*$ & 1.849±0.025 & 1:1 & 0.549 & 0.012 & 0.450 & 0.845 & 1486 & 1.00 & 0.07±0.01 & 0.42±0.06 & 0.51±0.08 & 0.05±0.01 & Dipolar & 1.20±0.24\\
TOI-2257 b	& 5.614±0.308$^*$ & 2.159±0.048 & 5:2 & 14.076 & 0.145 & 0.109 & 0.221 & 0.68 & 0.64 & 0.02±0.01 & 0.15±0.09 & 0.83±0.10 & 1.60±0.40 & Multip. & 0.01±0.002\\
TOI-237 b	& 2.686±0.156$^*$ & 1.483±0.046 & 1:1 & 5.436 & 0.034 & 0.066 & 0.133 & 3.6 & 0.62 & 0.12±0.01 & 0.72±0.07 & 0.16±0.08 & 0.71±0.07 & Multip. & 0.03±0.01\\
TOI-431 c	& 2.831±0.215 & 1.526±0.060 & 1:1 & 4.849 & 0.052 & 0.505 & 0.935 & 99.5 & 0.66 & 0.12±0.02 & 0.69±0.10 & 0.19±0.11 & 0.64±0.06	& Multip. & 0.03±0.01\\
TOI-451 b	& 4.374±0.252$^*$ & 1.910±0.070 & 1:1 & 1.859 & 0.029 & 0.759 & 1.388 & 778.9 & 1.00 & 0.06±0.02 & 0.36±0.14 & 0.58±0.16 & 0.19±0.02 & Multip. & 0.04±0.01\\
TOI-561 b	& 1.590±0.208 & 1.423±0.038 & 1:1 & 0.447 & 0.011 & 0.732 & 1.309 & 5167 & 1.00 & 0.07±0.03 & 0.44±0.15 & 0.49±0.17 & 0.04±0.002 & Dipolar & 0.41±0.08\\
TOI-561 d	& 11.962±0.735 & 2.529±0.075 & 1:1 & 25.620 & 0.157 & 0.710 & 1.309 & 23.4 & 0.99 & 0.05±0.02 & 0.27±0.12 & 0.68±0.14 & 4.44±0.44 & Multip. & 0.01±0.002\\
TOI-561 e	& 16.003±1.327 & 2.670±0.064 & 1:1 & 77.229 & 0.328 & 0.710 & 1.309 & 5.4 & 1.00 & 0.05±0.02 & 0.31±0.11 & 0.64±0.13 & 15.27±1.53 & Multip. & 0.01±0.002\\
TOI-763 b	& 9.793±0.452 & 2.280±0.064 & 1:1 & 5.606 & 0.060 & 0.748 & 1.380 & 177.8 & 1.00 & 0.08±0.02 & 0.45±0.10 & 0.47±0.11 & 0.79±0.08 & Multip. & 0.05±0.01\\
TOI-776 b	& 4.021±0.511 & 1.847±0.074 & 1:1 & 8.247 & 0.065 & 0.224 & 0.442 & 11.7 & 0.97 & 0.07±0.03 & 0.40±0.17 & 0.53±0.20 & 1.12±0.11	& Multip. & 0.01±0.002\\
TOI-776 c	& 5.489±0.966 & 2.008±0.078 & 1:1 & 15.665 & 0.100 & 0.222 & 0.442 & 5.0 & 0.88 & 0.07±0.03 & 0.39±0.18 & 0.54±0.21 & 2.40±0.24	& Multip. & 0.01±0.002\\
TOI-824 b	& 18.608±1.062 & 2.865±0.078 & 1:1 & 1.393 & 0.022 & 0.427 & 0.813 & 405.1 & 0.65 & 0.03±0.02 & 0.18±0.10 & 0.79±0.12 & 0.12±0.01 & Dipolar & 1.68±0.34\\
TOI-849 b	& 39.409±1.473 & 3.406±0.047 & 1:1 & 0.766 & 0.016 & 0.751 & 1.389 & 2503 & 0.55 & 0.02±0.01 & 0.09±0.05 & 0.89±0.06 & 0.04±0.01 & Dipolar & 1.29±0.52\\
pi Men c	& 4.855±0.471 & 2.040±0.028 & 1:1 & 6.268 & 0.068 & 1.087 & 1.968 & 307.3 & 0.94 & 0.04±0.02 & 0.22±0.10 & 0.74±0.12 & 0.73±0.07 & Multip. & 0.01±0.002\\

\hline
\end{longtable}

\tablefoot{$M_p$ is the planetary mass normalized to the Earth mass. Those values highlighted with an asterisk have been estimated following \citep{Chen17}. $R_p$ is the planetary radius normalized to the Earth radius. S:O shows the Spin:Orbit resonance. $P_{\rm rot}$  is the planetary rotational period in days. $a$, $D_2$ and $D_3$ are respectively are the semi-major axis, inner and outer boundaries of the HZ expressed in A.U. $S_{\rm eff}$ is normalized mean effective flux considering orbital eccentricity. {\it Prob.} It is the probability of being water-rich rocky planet. All the planets are tidally locked. Errors in the rotational period have not been included because they are so small that they have no impact on the magnetic properties of exoplanets.}
\end{landscape}
}
\end{appendix}

\end{document}